\documentclass[useAMS,usenatbib,usegraphicx]{mn2e}

\usepackage{amssymb}

\newcommand{\expect}{\mathbb{E}}
\newcommand{\disp}{\mathbb{D}}
\newcommand{\const}{\mathop{\rm const}\nolimits}

\newcommand{\FAP}{{\rm FAP}}
\newcommand{\EPC}{{\rm EPC}}
\newcommand{\Vol}{{\rm Vol}}
\title[Detecting non-sinusoidal periodicities]{Detecting non-sinusoidal periodicities in
observational data: the von Mises periodogram for variable stars and exoplanetary transits}
\author[R.V.~Baluev]{Roman V. Baluev\thanks{E-mail: roman@astro.spbu.ru}\\
Central Astronomical Observatory at Pulkovo of Russian Academy of Sciences, Pulkovskoje
shosse 65, St Petersburg 196140, Russia\\
Sobolev Astronomical Institute, St Petersburg State University, Universitetskij prospekt
28, Petrodvorets, St Petersburg 198504, Russia}

\begin{document}

\date{Accepted 2013 February 6.
      Received 2013 February 5;
      in original form 2012 December 26}

\pagerange{\pageref{firstpage}--\pageref{lastpage}} \pubyear{2013}

\maketitle

\label{firstpage}

\begin{abstract}
This paper introduces an extension of the linear least-squares (or Lomb-Scargle)
periodogram for the case when the model of the signal to be detected is non-sinusoidal and
depends on unknown parameters in a non-linear manner. The attention is paid to the problem
of estimating the statistical significance of candidate periodicities found using such
non-linear periodograms. This problem is related to the task of quantifying the
distributions of maximum values of these periodograms. Based on recent results in the
mathematical theory of extreme values of random field (the generalized Rice method), we
give a general approach to find handy analytic approximation for these distributions. This
approximation has the general form $e^{-z} P(\sqrt z)$, where $P$ is an algebraic
polynomial and $z$ being the periodogram maximum.

The general tools developed in this paper can be used in a wide variety of astronomical
applications, for instance in the studies of variable stars and extrasolar planets. For
this goal, we develop and consider in details the so-called von Mises periodogram: a
specialized non-linear periodogram where the signal is modelled by the von Mises periodic
function $\exp(\nu \cos \omega t)$. This simple function with an additional non-linear
parameter $\nu$ can model lightcurves of many astronomical objects that show periodic
photometric variability of different nature. We prove that our approach can be perfectly
applied to this non-linear periodogram.

We provide a package of auxiliary C++ programs, attached as the online-only material. They
should faciliate the use of the von Mises periodogram in practice.
\end{abstract}


\section{Introduction}
The \citet{Lomb76}-\citet{Scargle82} (hereafter LS) periodogram is a well-known powerful
tool, which is widely used to search for periodicities in observational data. The main
idea used in the LS periodogram is to perform a least-squares fit of the data with a
sinuous (harmonic) model of the signal and then to check how much the resulting value of
$\chi^2$ function improves for a given signal frequency. The maximum value of the LS
periodogram (i.e., the maximum decrement in the $\chi^2$ goodness-of-fit measure)
corresponds to the most likely frequency of the periodic signal. This natural idea is
quite easy to implement in numerical calculations. The linearity of the harmonic model
with respect to unknown parameters (two coefficients near the sine and cosine) introduces
additional simplifications.

Any signal detecting tool is not of much use without accompanying method of estimating the
statistical significance of candidate periodicities. Indeed, the random errors
contaminating the input data inspire noise fluctuations on the periodogram, so that we can
never be completely sure that the peak that we actually observed is a result of real
periodicity in the data. To assess the statistical significance of the observed
periodogram peak, we need to calculate the `false alarm probability' (hereafter $\FAP$)
associated with this peak. The $\FAP$ is the probability that the observed or larger
periodogram peak could be produced by random measurement errors. The smaller is $\FAP$,
the larger is the statistical significance. Given some small tolerance value $\FAP_*$
(say, $1\%$), we could claim that the detected candidate periodicity is statisticaly
significant (when $\FAP<\FAP_*$) or is not (when $\FAP>\FAP_*$).

From the statistical view point, the $\FAP$ is tightly connected with the probability
distributions of the periodogram considered under the null hypothesis (i.e., the
hypothesis of no signal in the data). If the frequency of the putative signal was known a
priori, we could use only single value of the LS periodogram to check whether the presence
of this periodicity is likely or not. In this case, the $\FAP$ is given by the well-known
exponential distribution of any single value $z$ of the LS periodogram, so that
$\FAP=e^{-z}$. However, the case in which the frequency of possible signal is basically
unknown is much more common. In this case, the $\FAP$ is provided by the distribution
function of the maximum value of the periodogram (corresponding to the frequency range
being scanned).

The calculation of the latter distribution is a non-trivial task. The absence of accurate
and/or rigorous analytic expression of this distribution (even for the plain LS
periodogram) represented a significant trouble for astronomers for about three decades. In
addition to the Lomb and Scargle works, it is worthwile to mention here the papers by
\citet{HorneBal86,Koen90,SchwCzerny98a,Cumming99,Cumming04,Frescura08}. We believe that
this obstacle was the main reason why basically no intricate extensions of the LS
periodogram attained enough practical popularity so far. Theoretically, it is not really
difficult to construct a periodogram where some fancy models of the data are used. Armed
with modern computers, we even may evaluate such periodograms in practice, even if they
rely on some CPU-greedy numerical algorithms. But what to do next? How to decide which of
the signals detected are real and which belong to the noise? The only general solution
available is the Monte Carlo simulation technique, which might be practically useful for
the basic LS periodogram, but not for more complicated cases, unfortunately.

Rather recently, a significant progress in this field was attained in the paper
\citep{Baluev08a}, where entirely analytic and simultaneously accurate approximations of
the $\FAP$ are given, based on the results in the theory of extreme values of stochastic
processes (the `Rice method'). In a brief form, the main result presented in
\citep{Baluev08a} for the LS periodogram is:
\begin{equation}
\FAP(z) \lesssim M(z) \approx W e^{-z} \sqrt z,
\label{ls_fap}
\end{equation}
where $z$ is the maximum periodogram value corresponding to a given frequency range, and
$W$ is the width of this range multiplied by a certain effective length of the time series
(which is usually close to the plain time span). The symbol `$\lesssim$' in~(\ref{ls_fap})
means that $\FAP(z)$ will never exceed $M(z)$ and simulataneously $M(z)$ represents an
asymptotic approximation for $\FAP(z)$, with the error decreasing for small $\FAP$ (or
large $z$). The high practical importance of the approximation in~(\ref{ls_fap}) is
founded on three things: (i) it is entirely analytic, eliminating any need for Monte Carlo
simulations, (ii) its practical accuracy usually appears good or at least quite
satifactory, and (iii) its possible errors never favour to more false alarms than we
expect, since we deal with an upper limit on $\FAP$.

The LS periodogram can be easily generalized in multiple ways to encompass more
complicated models \citep{SchwCzerny98a,SchwCzerny98b,Baluev08a,ZechKur09,FerrazMello81}.
First, we can introduce some base model of an expected underlying variation (typically a
long-term polynomial trend) and check whether the addition of a probe sinuos signal offers
enough improvement in $\chi^2$. These cases have been briefly considered in
\citep{Baluev08a} and our general conclusion was that such a modification does not
typically break the result~(\ref{ls_fap}). Second, we can deal with more complicated (but
still linear) models than just a sinusoid. In particular, in the work \citep{Baluev09a} we
considered the so-called multi-harmonic periodograms, where the periodic signal is
modelled by a trigonometric polynomial involving a few leading terms of the Fourier series
\citep{SchwCzerny96}. In this case, the formula~(\ref{ls_fap}) is generalized to
\begin{equation}
\FAP(z) \lesssim M(z) \approx W \alpha_n e^{-z} z^{n-1/2},
\label{multiharm_fap}
\end{equation}
where $\alpha_n$ are certain numbers depending on the degree $n$ of the approximating
trigonometric polynomial. Notice that $n=1$ corresponds to the LS case.

However, non-sinuous periodic signals, which are dealt with in astronomy, often obey
non-linear models. Then the use of the LS periodogram or periodograms from
\citep{Baluev08a,Baluev09a} is not optimal, since the corresponding periodic variation
might be fitted by an inadequate model. For instance, this is the case for lightcurves of
variable stars and for radial velocity curves of spectral binaries involving elongated
orbits. Theoretically, we could use a high-order Fourier expansion to approximate a
non-sinusoidal periodicity, but this solution is obviously inefficient due to an
unnecessarily large number of extra free parameters. The aim of the present paper is to
extend the results from \citep{Baluev08a} and \citep{Baluev09a} to the case of an
arbitrary model of the periodic signal, incorporating a few parameters in a non-linear
manner. As we will demonstrate, we can apply roughly the same technique (the Rice method)
to this case, with the major difference that we should now deal with random fields instead
of random processes. Namely, we will provide a closed approach to construct the limiting
approximation $M(z)$ in the form $W e^{-z} P(\sqrt z)$, where $P$ is an algebraic
polynomial.

The structure of the paper is as follows. In Section~\ref{sec_nonlindef}, we introduce a
general definition extending the LS periodogram to the non-linear case. In
Section~\ref{sec_FAP}, we consider the problem of assessing the statistical significance
of candidate periodicities detected with the non-linear periodogram. This description is
followed by an auxiliary Section~\ref{sec_coeffA} devoted to the ways of practical
evaluation of the theoretical approximations of Section~\ref{sec_FAP}. In
Section~\ref{sec_likper}, we discuss the concequences implied by various noise models of
the data. In Section~\ref{sec_examples}, a couple of concrete practical applications of
these results is supplied. In the first (rather tutorial) example, we aim to detect a
periodic signal of arbitrary (but a priori fixed) shape, when the unknown parameters are
the amplitude and the phase of the signal. In the second example, we consider a more
complicated periodogram based on the so-called von Mises model of the signal, essentially
$\exp(\nu\cos x)$, which involves an additional non-linear parameter $\nu$.

\section{Definition of the non-linear periodogram}
\label{sec_nonlindef}
Let $x_i$ denote the outcomes of $N$ observations made at timings $t_i$. The errors of
these measurements are assumed to follow Gaussian distributions and to be statistically
independent (hence, uncorrelated). The standard deviations of these errors, $\sigma_i$,
are assumed to be known a priori. We want to test, whether these observations are
consistent with some base model of variation, or certain deterministic periodicity is also
present.

The data model to be tested for consistency with the data is $\mu_{\mathcal
H}(t,\btheta_{\mathcal H})$, where the vector $\btheta_{\mathcal H}$ incorporates
$d_{\mathcal H}$ unknown parameters, which should be estimated from the data. We assume
that this model is linear with respect to unknown parameters:
\begin{equation}
\mu_{\mathcal H}(t,\btheta_{\mathcal H}) = \btheta_{\mathcal H}\cdot \bvarphi_{\mathcal H}(t),
\label{muH}
\end{equation}
where the vector of base functions $\bvarphi_{\mathcal H}(t)$ is set a priori. Typically,
the model $\mu_{\mathcal H}$ incorporates a free constant term, and, possibly, a long-term
polynomial trend with free coefficients. The Lomb-Scargle periodogram, by the way, assumes
that $\mu_{\mathcal H}\equiv 0$, implicitly requsting some preliminary centering of the
time series.

The model of the periodic signal is given by $\mu(t,\btheta,f)$, where the vector
$\btheta$ contains $d$ unknown parameters to be estimated from the data together with the
frequency $f$. The united vector $\btheta_{\mathcal K} = \{\btheta_{\mathcal H},\btheta\}$
parametrizes the compound alternative model of the data\footnote{We use here all notation
conventions used in \citep{Baluev08a}. For instance, the braces $\{*,*,\ldots\}$ denote
the association of vectorial or scalar arguments into a single vector, the angular
brackets $\langle * \rangle$ denote the summation of the argument over timings $t_i$ with
weights $1/\sigma_i^2$, and $\bmath x \otimes \bmath y \equiv \bmath x \bmath y^{\rm T}$
is the dyadic product of the vectors $\bmath x$ and $\bmath y$.}, which is given by
\begin{equation}
\mu_{\mathcal K}(t,\btheta_{\mathcal K},f) = \mu_{\mathcal H}(t,\btheta_{\mathcal H}) + \mu(t,\btheta,f).
\label{muK}
\end{equation}

We denote the $d$-dimensional domain, where $\btheta$ is supposed to reside, by $\Theta$.
The signal is supposed to vanish when $\btheta$ belongs to some `null domain' $\Theta_0
\subset \Theta$ and not to vanish when $\btheta$ does not belong to $\Theta_0$. Therefore,
we wish to test, whether the data are consistent with the base hypothesis $\mathcal H:
\btheta\in\Theta_0$ (implying that the model $\mu_{\mathcal H}(t,\btheta_{\mathcal H})$
fits the data satisfactory) or this base hypothesis should be rejected in favour of the
alternative $\mathcal K: \btheta \in \Theta\setminus\Theta_0$ (implying the model
$\mu_{\mathcal K}(t,\btheta_{\mathcal K},f)$). The model $\mu$ may be non-linear with
respect to $\btheta$.

The unknowns $\btheta_{\mathcal H}, \btheta$, and $f$ can be estimated using the
least-squares approach. Under the hypothesis $\mathcal H$, the best-fitting estimation of
$\btheta_{\mathcal H}$ can be obtained in result of minimizing the functions
$\chi_{\mathcal H}^2(\btheta_{\mathcal H}) = \langle (x-\mu_{\mathcal H})^2 \rangle$.
Under the hypothesis $\mathcal K$, the best-fitting estimations of $\btheta_{\mathcal H},
\btheta$ and $f$ should correspond to the minimum value of the function $\chi_{\mathcal
K}^2(\btheta_{\mathcal K},f) = \langle (x-\mu_{\mathcal K})^2 \rangle$. Since
$\mu_{\mathcal H}$ is linear, the minimizations by $\btheta_{\mathcal H}$ can be performed
rapidly and precisely using the usual linear least-squares algorithms. The minimization of
$\chi^2_{\mathcal K}$ over the remaining variables is equivalent to the maximization of
the non-linear function
\begin{equation}
\zeta(\btheta,f) = \frac{1}{2} \left[\min_{\btheta_{\mathcal H}} \chi^2_{\mathcal H}(\btheta_{\mathcal
H}) - \min_{\btheta_{\mathcal H}}\chi^2_{\mathcal K}(\btheta_{\mathcal H},\btheta,f)\right],
\label{nonlinprdg}
\end{equation}
which simultaneously characterises the improvement in the $\chi^2$ fit quality, which is
achieved by means of adding to the base model the model of the periodic signal with given
values of $\btheta$ and $f$. Note that the maxima of $\zeta(\btheta,f)$ do not depend on
the choice of the parametrization. That is, they are invariable with respect to a
non-degenerated transformation of the vector $\btheta$ (and any non-degenerated linear
transformation of $\btheta_{\mathcal H}$).

We can perform the minimization over the frequency $f$ in a traditional manner by means of
looking for the highest peak on the graph of the function
\begin{equation}
z(f) = \max_{\btheta \in \Theta} \zeta(\btheta,f),
\end{equation}
which may be called the ``non-linear least-squares periodogram''. This definition means
that for any fixed frequency $f$ we perform the fit of our model via the remaining $d$
parameters $\btheta$. The value of the periodogram characterizes the relevant advance in
the $\chi^2$ fit quality. When the model $\mu(t,\btheta,f)$ is linear with respect to
$\btheta$, this definition of $z(f)$ coincides with the definition of the linear
least-squares periodogram from \citep{Baluev08a}.

In the majority of practical applications, one of the parameters in $\btheta$ is the
amplitude $K$ of the periodic variation. This means that
\begin{equation}
 \mu(t,\btheta,f) = K h(t,\bxi,f),
\label{ampl}
\end{equation}
where the vector $\bxi$ contains $d-1$ remaining unknown parameters of the signal. We
assume that $\bxi$ belongs to some domain $\Xi$ in $d-1$ dimensions, so that the domain
$\Theta$ represents the Cartesian product $[0,+\infty) \times \Xi$ or $(-\infty,+\infty)
\times \Xi$, and the null domain is the domain of zero amplitude: $\Theta_0 = \{K=0\}
\times \Xi$. For simplicity, let us firstly consider the case when $d_{\mathcal H}=0$ and
the hypothesis $\mathcal H$ states that the data do not contain anything but the white
Gaussian noise. In this case, $\chi^2_{\mathcal H}\equiv \langle x^2 \rangle$ and
$\zeta(\btheta,f) = \langle x \mu \rangle - \langle \mu^2 \rangle/2 = \langle x h \rangle
K - \langle h^2 \rangle K^2/2$ is a quadratic polynomial of $K$, which can be easily
maximized given fixed $f$ and $\bxi$. This results in a least-squares estimation $K^* =
\langle x h \rangle / \langle h^2\rangle$, and in the maximum $\max_K \zeta = \eta^2/2$,
where $\eta(\bxi,f)$ represents the new function to be maximized by the remaining
parameters. It can be expressed as
\begin{equation}
 \eta(\bxi,f) = \langle x \psi\rangle,
\label{eta}
\end{equation}
where $\psi(t,\bxi,f) = h(t,\bxi,f)/\sqrt{\langle h^2 \rangle}$.

A similar result may be obtained for the case when the relation~(\ref{ampl}) is still
valid, but the model $\mu_{\mathcal H}$ is no longer empty. It is not hard to check that,
if the models $\mu$ and $\mu_{\mathcal H}$ were orthogonal in the sense that $\langle h
\bvarphi_{\mathcal H} \rangle = 0$, the maximum of $\zeta$ by $K$ could be expressed
exactly in the same way as it was described in the previous paragraph. In the general case
the models are not orthogonal, and we introduce the new model function
\begin{equation}
\tilde h(t,\bxi,f) = h(t,\bxi,f) - ( \mathbfss Q_{\btheta_{\mathcal H},\btheta_{\mathcal
H}}^{-1} \mathbfss Q_{\btheta_{\mathcal H},K}) \cdot \bvarphi_{\mathcal H}(t),
\end{equation}
where $\mathbfss Q_{\btheta_{\mathcal H},\btheta_{\mathcal H}} = \langle
\bvarphi_{\mathcal H} \otimes \bvarphi_{\mathcal H}\rangle$ is the $d_{\mathcal H}\times
d_{\mathcal H}$ Fisher information matrix associated with $\btheta_{\mathcal H}$, and
$\mathbfss Q_{\btheta_{\mathcal H},K} = \langle \bvarphi_{\mathcal H} h\rangle$ is the
$d_{\mathcal H}\times 1$ Fisher information matrix for the parameters $\btheta_{\mathcal
H}$ and $K$. Since the identity $\langle \tilde h \bvarphi_{\mathcal H} \rangle = 0$ holds
true, the new model of the signal is orthogonal to the base model. Now $\tilde h$ should
replace $h$ in the expression for $\psi$, so that
\begin{equation}
\psi(t,\bxi,f) = \frac{h(t,\bxi,f) - (\mathbfss Q_{\btheta_{\mathcal H},\btheta_{\mathcal
H}}^{-1} \mathbfss Q_{\btheta_{\mathcal H},K}) \cdot \bvarphi_{\mathcal H}(t)}
{\sqrt{\langle h^2\rangle - \mathbfss Q_{K,\btheta_{\mathcal H}} \mathbfss
Q_{\btheta_{\mathcal H},\btheta_{\mathcal H}}^{-1}
\mathbfss Q_{\btheta_{\mathcal H},K}}}.
\label{psi}
\end{equation}
After that, we can directly calculate the quantity $\eta$ from the equation~(\ref{eta}). Finally,
\begin{equation}
\max_K \zeta = \eta^2/2
\label{maxKzeta}
\end{equation}
with
\begin{equation}
K^* = \eta \left/
\sqrt{\langle h^2\rangle - \mathbfss Q_{K,\btheta_{\mathcal H}} \mathbfss
Q_{\btheta_{\mathcal H},\btheta_{\mathcal H}}^{-1} \mathbfss Q_{\btheta_{\mathcal H},K}}\right. .
\label{bestfitK}
\end{equation}
The best-fitting values of $\bxi$ and $f$ correspond to the maximum of $\eta$.

Often it might be useful to assume that negative values for $K$ are not allowed. Then we
should make a small amendment to the last formulae~(\ref{maxKzeta}) and~(\ref{bestfitK}).
Namely, they can be used only for $\eta>0$, while for $\eta<0$ we should set $\max_K \zeta
= 0$ and $K^*=0$ with the best-fitting values of $\bxi$ and $f$ undefined.

The formulae become more simple for an important practical case when $d_{\mathcal H}=1$
and $\varphi_{\mathcal H}\equiv 1$ reflects a free constant offset of the data. In this
case, let us first define
\begin{equation}
x_c = \langle x \rangle / \langle 1 \rangle, \quad h_c(\bxi,f) = \langle h \rangle / \langle 1 \rangle,
\quad D = \langle (h-h_c)^2 \rangle,
\label{flm1}
\end{equation}
and then evaluate
\begin{equation}
\eta = \langle (x-x_c) (h-h_c) \rangle / \sqrt D, \quad \max_K \zeta=\eta^2/2, \quad K=\eta/\sqrt D.
\label{flm2}
\end{equation}
Note that the quantity $\langle 1 \rangle$ represents the sum of weights of all
observations.

\section{Approximating the false alarm probability using the Rice method}
\label{sec_FAP}
\subsection{General introduction to the problem}
In this paper we are interested in the false alarm probability ($\FAP$) associated with
the observed maximum peak $z_{\rm max} = \max_{0\leq f\leq f_{\rm max}} z(f)$, where
$f_{\rm max}$ is some a priori given maximum frequency. This false alarm probability can
be formally defined as follows:
\begin{equation}
\FAP(z_{\rm max}) = \Pr\{\exists \btheta \in \Theta, f\in [0,f_{\rm max}]: \zeta(\btheta, f)>z_{\rm max} \},
\label{FAPmax}
\end{equation}
again with the probability operator calculated under the null hypothesis (no actual signal
in the data). We can see that to assess the $\FAP$, we should know the distribution
function of the maximum values of $\zeta(\btheta,f)$. This function represents a
real-valued random field defined on a domain of dimension $d+1$.

One could claim that since the model of the extra signal contains $d$ free parameters and
since the quantity $z(f)$ is the logarithm of the likelihood ratio statistic, the
distribution of $2z(f)$ (for a fixed $f$) should tend to the $\chi^2$ distribution with
$d$ degrees of freedom, when the number of observations grows. This was assumed, for
instance, by \citet{Cumming04} who considered the case of Keplerian velocity variation
with four unknown parameters (plus the period). We must caution the reader that in general
this assumption is incorrect because the conditions of the corresponding limiting theorem
are not satisfied. The most important reason comes from the fact that the parameters
$\bxi$ have no physical sense (are undefined) when $K=0$. Speaking mathematically, they
are not identifiable for $K=0$. The lack of identifiability under the null hypothesis
usually destroys the usual asymptotic properties of the likelihood ratio (and $\chi^2$)
tests \citep{DacunhaGassiat99}. This is because we typically just cannot construct a valid
Taylor series of the signal model at $K=0$, except for rare special cases. Without that we
cannot linearize this model under the null hypothesis, which is critical for the validity
of the asymtotic $\chi^2$ distribution. An exception is provided, for instance, by the LS
periodogram with the harmonic model of the signal. In this special case, we are able to
perform the following re-parametrization: $K\cos(2\pi f t + \lambda) = a \cos(2\pi f t) +
b \sin(2\pi f t)$. While that phase $\lambda$ was a not identifiable at $K=0$, the new
parameters $a$ and $b$ are already identifiable and even linear. In this case, the
distribution of each single value of $2z(f)$ is indeed the $\chi^2$ one with two degrees
of freedom. Unfortunately any similar trick is not possible for the majority of the other
models, even apparently simple ones.

The things get even worse for the more practical case when $f$ is also unknown. In this
case the frequency, trated as a new free parameter, is not identifiable (at $K=0$) even
for the LS periodogram. Actually, we may note that non-identifiability of the frequency is
the primary obstacle that made the treatment of the significance levels of the LS
periodogram so difficult and non-rigorous over decades. The LS periodogram of the noise
containes an infinite sequence of similar noisy narrow peaks, but none of them can serve
as a reference position for a quadratic Taylor approximation that would be valid in the
whole frequency range. If not that obstacle then we could just use the $chi^2$
distribution with three degrees of freedom (two for $a$ and $b$ plus one for $f$) to
approximate the necessary distribution of the LS periodogram. However, in
\citep{Baluev08a} we managed to deal with this obstacle using the so-called `Rice method',
treating the noisy LS periodogram as a random process depending on a real argument $f$,
which was a single non-linear parameter of the model. The case of non-linear periodograms
just adds more non-linear arguments of $\zeta$, but the issue of their non-identifiability
at $K=0$ remains qualitatively the same. Therefore, we may try to treat this more general
situation using the same or similar method.

Of course, it is hardly possible to derive an exact expression for $\FAP$, but we would be
pretty satisfied if we find at least an approximation analogous to what we obtained in our
previous works. Namely, we aim to obtain something like
\begin{equation}
\FAP(z_{\rm max}) \lesssim M(z_{\rm max}),
\label{bound_fap}
\end{equation}
where $M$ represents simultaneously an upper bound for $\FAP$ and its more or less good
asymptotic approximation for large $z_{\rm max}$ (small $\FAP$).

\subsection{Basic ideas of the Rice method}
The modern comprehensive theory of the Rice method and relevant topics can be found in the
reviews \citep{Kratz,AzaisWschebor-levelsets}. Here we present only a very brief
extraction of the results that are most useful in our present paper. Suppose we deal with
some arbitrary random process or field $Z(\bmath x)$ and we need to find the probability
that its maximum (within some domain $\bmath x \in \mathbb X$) will lie beyond a specified
level $Z(\bmath x)=z$. In our signal detection task this probability is equal to
$\FAP(z)$, and this is obviously a complementary probability to the distribution function
of the maximum of $Z$. The general Rice method to estimate these thing is based on two
main points. First, we should construct some derived integer random variable $\mathcal
N(z)$, such that the event $\mathcal N(z)=0$ is equivalent (or almost equivalent) to the
event $\{Z(\bmath x)<z\, \forall \bmath x \in \mathbb X\}$, and the event $\mathcal
N(z)\geq 1$ is (almost) equivalent to $\{\exists \bmath x \in \mathbb X: Z(\bmath x)>z\}$.
The boundary event when $Z(\bmath x)\leq z$ everywhere in $\mathbb X$ and there is only
one or a few $\bmath x$ such that $Z(\bmath x)=z$ should usually correspond to $\mathcal
N(z)=1$. The word ``almost'' refers here only to the effects at the boundary of $\mathbb
X$ (the boundary maxima); if we somehow knew for sure that boundary maxima are impossible
than this word can be just omitted.

For random processes a good choice for $\mathcal N$ is the number of up-crossings of the
specified level; i.e. the number of points $x$ such that $Z(x)=z$ and $Z'(x)>0$. For
random fields the term of up-crossing is meaningless, and in this case we choose $\mathcal
N(z)$ to be the number of local maxima beyond $z$ (and inside $\mathbb X$), that is the
number of points $\bmath x$ where $Z>z$, $Z'=0$, and $Z''$ is negative-definite.

Given such a counter variable $\mathcal N$, we can estimate the required false alarm
probability, i.e. the probability for $Z(\bmath x)$ to exceed a given level $z$ somewhere
in $\mathbb X$, as
\begin{eqnarray}
\FAP(z) \lesssim M(z) = M_{\rm boundary}(z) + \tau(z), \nonumber\\
\tau(z) = \expect\mathcal N(z).
\label{FAPbound}
\end{eqnarray}
Here the term $M_{\rm boundary}$ refers to the maxima attained at the boundary of $\mathbb
X$; it may or may not be neglected, depending on other conditions of the task. We will
discuss it in detail later. The primary term is $\tau(z)$, which is equal to the
mathematical expectation of the selected counter. This formula is basically the same
as~(\ref{bound_fap}) with concretized function $M(z)$.

The second point of the Rice method is the generalized Rice formula for $\tau$. For the
random processes we have
\begin{equation}
\tau(z) = \int\limits_{\mathbb X} \expect([Z'(x)]_+ \mid Z(x)=z)\, p_Z(z)\, dx,
\label{Ricefrm_proc}
\end{equation}
where the functions $p$ stand for the probability density functions of the quantity $Z(x)$
shown as indices (this $p$ also depends on $x$), and $[a]_+=\max(0,a)$. If necessary, the
expression~(\ref{Ricefrm_proc}) can be obviously rewritten in terms of the joint
distribution of $Z(x)$ and $Z'(x)$, see \citep{Baluev08a}. The name ``Rice method'' and
``Rice formula'' are after \citet{Rice44}, who constructed his original Rice formula for
the stationary Gaussian random process.

In this paper we will use the generalized Rice formula for random fields. Actually, now we
have even three formulae of that type. The first one is introduced in
\citep{AzaisDelmas02}; it can be written down as
\begin{equation}
\tau(z) = \int\limits_z^{\infty} dZ \int\limits_{\mathbb X} \expect(\det [Z'']_- \mid Z'=0; Z)\,
 p_{Z,Z'}(Z,0)\, d\bmath x,
\label{Ricefrm_field}
\end{equation}
where $\det[Z'']_-$ is equal to $|\det Z''|$ when $Z''$ is negative-definite, and zero
otherwise. This formula is generally similar to~(\ref{Ricefrm_proc}), although
considerably more complicated. \citet{AzaisWschebor-levelsets} introduced in their
Chapter~8 a variation of~(\ref{Ricefrm_field}) with $\det [Z'']_-$ replaced by $|\det
Z''|$. Such replacement obviously somewhat increases the right-hand side
of~(\ref{Ricefrm_field}), keeping its upper-limit property, but making the computations a
bit more easy. It counts \emph{all} the critical points of $Z(\bmath x)$ above $z$, not
just the local maxima. However, both these formulae are usually too difficult for
computations, and the formula that is typically used in practice contains just the
``naked'' $\det Z''$ instead of $|\det Z''|$ or $\det[Z'']_-$. Such a formula gives the
mathematical expectation of the Euler-Poincar{\'e} characteristic ($\EPC$) of the
level-section set $\{\bmath x\in \mathbb X: Z(\bmath x)>z\}$ (which is also called
as the ``excursion set'').

Unfortunately, the quantity $\expect(\EPC)$ does not strictly retain the upper-limit
property of $\expect \mathcal N$. However, it is known (at least for Gaussian fields, see
e.g. Chapter~8 by \citet{AzaisWschebor-levelsets}) that for large levels $z$ the
quantities $\expect \mathcal N$ and $\expect(\EPC)$ are asymptotically equivalent, and
their difference decreases rather quickly (we will detail this below). This is because
beyond a large $z$ all critical points of $Z$ are local maxima with almost unit
probability, so the relevant excursion set represents a number of (filled) ellipsoids
encompassing the positions of these maxima. Each such a filled ellipsoid has $\EPC=1$, and
thus $\EPC\simeq \mathcal N$ for large $z$. All this means that we can typically use
$\expect(\EPC)$ as a good practical approximation for $\tau$. Even if $\expect(\EPC)$ does
not provide an entirely strict upper bound, the relevant errors usually appear negligible
for practical levels of $z$. Of course we must admit that ``usually'' or ``typically'' is
not the same as ``always'', but nonetheless this approximation appears quite satisfactory
in the examples considered below in the paper, as well as in a few other cases that we
prepare for a future publication.

The Rice method usually provides good practical accuracy, so that the mentioned upper
bound~(\ref{bound_fap}) appears close to the actual value of $\FAP$, at least for
practically important case of small $\FAP$ levels. The Rice method does not belong to
widely-known methods, because it is not mentioned in a typical handbook on mathematical
statistics. Therefore, its usage in applications (e.g. in astronomy) is rare. However,
rare does not mean absent: we found that some variant of this method was applied by
\citet{Bardeen86} to study cosmological density fluctuations, which were modelled by a
Gaussian random field.

\subsection{Applying the Rice method to non-linear periodograms}
Mathematically, the condition of $\max\zeta \leq z$ is equivalent to that of $\max|\eta|
\leq \sqrt{2z}$ (case of arbitrary $K$) or $\max\eta \leq \sqrt{2z}$ (case of $K\geq 0$).
Therefore, we need to calculate the distribution of the maximum values of the random
function $\eta(\bxi,f)$ to estimate the $\FAP$. In this subsection we limit ourself by the
single-sided case $K\geq 0$, bearing in mind that to obtain the formulae for the case of
arbitrary $K$ we need to double the right-hand side of~(\ref{FAPbound}), because then we
need to honour the maxima of $\eta$ above $\sqrt{2z}$ as well as its minima below
$-\sqrt{2z}$, which are entirely analogous.

The random field $\eta(\bxi,f)$ possesses quite simple statistical properties. From the
definition~(\ref{eta}) it clearly follows that if the noise in our observations is
Gaussian, $\eta$ represents a Gaussian random field. It is easy to check from~(\ref{eta})
and~(\ref{psi}) that $\expect\eta\equiv 0$ and the variance $\disp\eta\equiv 1$. This
places us in the framework of Theorem~1 by \citet{AzaisDelmas02}. In this case we can
use~(\ref{FAPbound}) with
\begin{equation}
\tau(z) \approx \expect(\EPC)= \sum_{j=0}^{[n/2]} a_j P_{n+1-2j}(z).
\label{taunonlin}
\end{equation}
In this relation, the integer $n$ represents the number of free arguments of $\eta$. It is
equal to $\dim\bxi=d-1$ or $\dim\bxi+1 = d$, depending on whether we consider the case of
fixed or free frequency $f$. Below we will only consider the more practical second case
with $n=d$, but in order to avoid misunderstandings we prefer to keep different notations
for $n$ and $d$. The notation $[*]$ stands for the integer part of the argument. The
functions $P_k(z)$ represent the tail probability associated with the $\chi^2$
distribution with $k$ degrees of freedom:
\begin{equation}
P_k(z) = \frac{1}{\Gamma(k/2)} \int\limits_z^\infty x^{k/2-1} e^{-x} dx.
\end{equation}

When $z\to\infty$, the error of the approximation in~(\ref{taunonlin}) has the decrease
rate quicker than $e^{-(1+\delta)z}$ with some positive $\delta$, while $\tau$ itself
typically decreases as $e^{-z} z^{(n-1)/2}$. This means that the relevant \emph{relative}
error decreases qucker than $e^{-z\delta}/z^{(n-1)/2}$.

We can see that the expression~(\ref{FAPbound}) involves a linear combination of $\chi^2$
tails with different numbers of degrees of freedom. However, the sum of the coefficients
$a_j$ is not necessary unit, so that the expansion in~(\ref{FAPbound}) is not a mixture of
distributions in the rigorous meaning of this notion. For large $z$, the first term with
$P_{n+1}\sim z^{(n-1)/2} e^{-z}$ dominates, whereas the relative magnitudes of the
remaining terms decrease as $\sim 1/z^j$.

The calculation of the coefficients $a_j$ represents a large technical difficulty. These
coefficients are proportional to the quantities $k_{2j}$ introduced in Theorem~1 by
\citet{AzaisDelmas02}. We altered the original coefficients $k_{2j}$ in order to have the
functions $P_k$ explicitly in the sum~(\ref{taunonlin}).

Let us denote the variance-covariance matrix of the gradient of $\eta$ as $\mathbfss G$
(it is denoted as $\Lambda$ by \citet{AzaisDelmas02}). It can be easily calculated from
the eq.~(\ref{eta}). The part of the matrix $\mathbfss G$ corresponding to only the
parameters $\bxi$ is equal to
\begin{equation}
\left\langle \frac{\partial \psi}{\partial \bxi} \otimes \frac{\partial
         \psi}{\partial \bxi} \right\rangle,
\label{gradvar}
\end{equation}
and the remaining elements due to the frequency parameter can be expressed in an entirely
analogous manner.

At first, let us consider a simplified situation when $\mathbfss G$ does not depend on
$\bxi$ and $f$. Then we can use the proposition ``a'' of Theorem~1 in
\citep{AzaisDelmas02}. We find that in this case the coeffiecients $a_j$ (or $k_{2j}$) are
proportional to the coefficients of the Hermite polynomials $H_n$ \citep[\S 21.7]{Korn},
so that after some elementary transformations we have
\begin{equation}
\tau \approx A F_n(z)
\label{FAPhermite}
\end{equation}
with
\begin{eqnarray}
A &=& \frac{\sqrt{\det\mathbfss G}}{2\pi^{(n+1)/2}} \Vol(\Xi) f_{\rm max}, \nonumber\\
F_n(z) &=& \int\limits_z^\infty H_n(\sqrt x) e^{-x} \frac{dx}{\sqrt x} = e^{-z} H_{n-1}\left(\sqrt z\right).
\label{coefAconst}
\end{eqnarray}
The last equality in~(\ref{coefAconst}) can be just checked by direct differentiation or
it can be derived ``honestly'' using the Rodrigues representation for $H_n$. Notice that
$H_n$ are normalised here so that their highest coefficients are equal to unit and the
weighting function is $e^{-x^2}$.

\begin{figure}
\includegraphics[width=84mm]{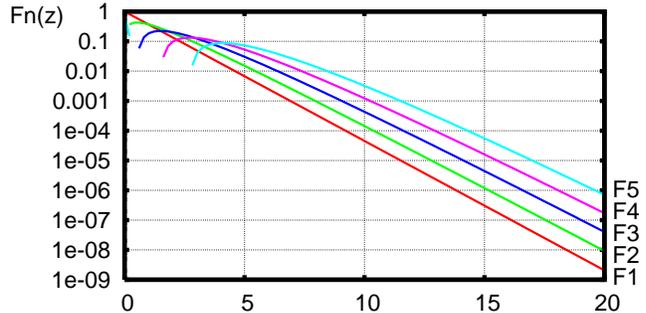}
\caption{The graphs of several functions $F_n(z)$.}
\label{fig_funcFn}
\end{figure}

Let us write down a few of the functions $F_n$:
\begin{eqnarray}
F_1(z)=e^{-z},\quad F_2(z)=e^{-z}\sqrt{z},\, F_3(z)=e^{-z}\left(z-\frac{1}{2}\right),\nonumber\\
F_4(z)=e^{-z}\left(z-\frac{3}{2}\right)\sqrt z,\, F_5(z)=e^{-z}\left(z^2-3z+\frac{3}{4}\right).
\end{eqnarray}
We plot the graphs of these functions in Fig.~\ref{fig_funcFn}. Also, a notable recursive
relation $F_{n+1}(z) = -F_n'(z) \sqrt z$ can be used.

When $\mathbfss G$ is non-constant, which is a more practical case, we can easily
calculate only the primary term in~(\ref{taunonlin}). We have in this case
\begin{equation}
\tau \approx A e^{-z} z^{(n-1)/2}
\label{coefAnonconst}
\end{equation}
with the relative error of $\sim 1/z$, which is worse than the error of~(\ref{taunonlin}).
Here
\begin{equation}
A = \frac{1}{2\pi^{(n+1)/2}} \int\limits_0^{f_{\rm max}} df \int\limits_\Xi
\sqrt{\det\mathbfss G}\, d\bxi,
\label{coefA}
\end{equation}
which is an evident generalization of $A$ from~(\ref{coefAconst}).

The remaining $z$-power terms are different from the case of constant $\mathbfss G$, and
usually they are very hard to evaluate, because they involve conditional covariances of
the \emph{second}-order derivatives of $\eta$ in quite unpleasant combinations (see
proposition ``b'' of Theorem~1 by \citealt{AzaisDelmas02}). It might be noticed that for a
simple case $n=1$ we have only one term in~(\ref{taunonlin}).

\subsection{The role of the boundary effects}
The quantity $\tau$ on itself does not yet provide a closed solution of the problem.
According to~(\ref{FAPbound}), we need to assess a similar term, which is related to the
number of the so-called ``boundary maxima'' (which are the local maxima of the random
field restricted to the domain $\mathbb X$ boundary).

This term takes into account the situation when all local maxima in the domain interior
appear smaller than some boundary maximum. It can be calculated using very similarly to
the term $\tau$, we just need to consider the restriction of our task to this boundary and
apply the Rice method in a recursive manner. We should approximate the quantity $M_{\rm
boundary}$ by a formula similar to~(\ref{FAPbound}) and~(\ref{taunonlin}), but with $n$
replaced by $n-1$. Hence, its relative contribution decreases for large $z$, but at a
rather slow rate of $\sim 1/\sqrt z$.

We however must take care of one small but rather important thing. When we restrict our
field $\eta$ to the boundary of our domain, a maximum at the boundary does not necessarily
represent a good candidate for the global maximum. Whether a particular boundary maximum
is ``good'' or ``bad'', depends on the sign of the derivative of $\eta$ in the direction
of the outward normal to the boundary (i.e., the projection of the gradient $\eta'$ to
this normal). If it is negative, we can definitely find larger values of $\eta$ when
stepping from the boundary inwards. Thus, such a boundary maximum can never provide the
global maximum of $\eta$, so we call it ``bad''. When counting the boundary maxima, we
must filter out all ``bad'' ones, keeping only the ``good'', which offer positive outward
derivative. Mathematically, all this requests from us to replace in~(\ref{Ricefrm_field})
the gradient $Z'$ and the Hessian $Z''$ by their projections to the tangent plane to the
boundary, (denote them, say, $Z'_\parallel$ and $Z''_\parallel$), and consider the
operator $\expect$ and the probability density $p_{Z,Z'_\parallel}$ conditionally to an
additional constraint $Z'_\perp>0$ (where index $\perp$ means the projection on the
outward normal to the boundary surface).

In practice this usually just means that we need to halve (precisely or approximately) the
estimated number of the boundary maxima, $\tau_{\rm boundary}$. We must admit that this
issue has not been investigated in the literature with enough details.
\citet{AzaisDelmas02} prove this ``$1/2$-rule'' under certain restrictive
assumptions, among which the most important is the requirement of constant $\mathbfss G$.
From their Theorem~3 it follows, basically, that
\begin{equation}
M_{\rm boundary} = \frac{1}{2} \tau_{\rm boundary} + \ldots,
\label{Mboundary}
\end{equation}
where $\tau_{\rm boundary}$ can be evaluated in essentially the same manner as $\tau$,
considering the restriction of the task to the boundary surface (which implies, in
particular, a decrease in $n$), and under ``$\ldots$'' we mean here some terms having
faster decrease rate than $\tau_{\rm boundary}$. Although \citet{AzaisDelmas02} leave the
case when $\mathbfss G \neq \const$ aside, after investigation of their detailed proofs,
we do not find an obstacle in generalizing their single-term asymptotic formula for
$M_{\rm boundary}$ to a more general case with $\mathbfss G \neq \const$. Moreover, we
find that the main neglected term contained in ``$\ldots$'' of~(\ref{Mboundary}) has the
relative magnitude of $\sim 1/z$.

Since in the general case we anyway keep only the greatest terms in $\tau$ and similar
quantities, we can use a two-term approximation like
\begin{equation}
\FAP(z) \lesssim M(z) \simeq \left( A z^{(n-1)/2} + \frac{1}{2} A_{\rm boundary} z^{n/2-1} \right) e^{-z}.
\end{equation}
The principal error of the right-most expression has the relative magnitude of $\sim 1/z$
and is due to the omitted terms in $\tau$, while the omitted terms in $\tau_{\rm
boundary}$ are of even a smaller order $\sim 1/z^{3/2}$.

In more complicated cases, the boundary itself may be non-smooth due to ``sub-boundaries''
of smaller dimension (edges, vertexes), which will generate extra terms in $M_{\rm
boundary}$. It is rather difficult to formulate a simple and general receipt of how to
deal them with, because the geometry of the boundary might be quite complicated in
general. However, later we explain this procedure on a concrete example of the von Mises
periodogram, when the parametric domain is a rectangle.

\section{Evaluating the coefficient $A$}
\label{sec_coeffA}

\subsection{Assuming the uniform phase coverage}
\label{subsec_Aavg}
It would be useful to construct the analytic expressions of the $\FAP$ for the case when
our observations are distributed approximately uniformly in time. We assume that the
timings $t_i$, when they are phased to some frequency $f$, cover the relevant phase more
or less uniformly. In this case, we can
approximate the summation $\langle * \rangle$ over the time series by means of integration
over its time span $[t_1,t_2]$. Saying it more accurately, we approximate the time-series
average $\langle * \rangle/\langle 1\rangle$ by the integral average over the time span.
Moreover, the periodic character of the model $\mu$ usually allows for this integration to
be performed over a single period only. The periodicity of $h$ implies that $h(t,\bxi,f) =
g(2\pi f t + \lambda, \bnu)$, where the function $g(x,\bnu)$ is $2\pi$-periodic in $x$ and
$\lambda$ is the phase parameter. The remaining $d-2$ parameters form the vector $\bnu$.
This means that $\Xi = [0,2\pi] \times \Upsilon$ and $\Theta = [0,+\infty) \times [0,2\pi]
\times \Upsilon$, where $\Upsilon$ is some $(d-2)$-dimensional domain of parameters
$\bnu$. Now we can approximate, for instance, the mean value of $h$ as
\begin{equation}
\left\langle h \right\rangle \approx \frac{w}{P} \int\limits_0^P h(t,\bxi,f) dt  =
\frac{w}{2\pi} \int\limits_0^{2\pi} g(x,\bnu) dx = w \overline{g},
\end{equation}
where $P=1/f$ is the period of the signal, $w=\langle 1 \rangle$ is the sum of the weights
and the overline denotes the continuous averaging over the periodic variable $x$. Note
that usually we deal with the case when the base model $\mu_{\mathcal H}$ incorporates a
constant term. According to~(\ref{psi}), this means that we should first subtract the
constant $\langle g\rangle/w \approx \overline{g}$ from the model of the signal. We will
assume that $g$ was already centered.

The same arguments lead to the equality $\langle h^2 \rangle \approx w \overline{g^2}$.
The latter expression does not depend on $f$ and $\lambda$, so we can write down the
derivatives of $\psi$ as
\begin{equation}
\frac{\partial \psi}{\partial f} \approx \frac{2\pi t\,g'_x}{\sqrt{w \overline{g^2}}}, \,
\frac{\partial \psi}{\partial \lambda} \approx \frac{g'_x}{\sqrt{w \overline{g^2}}}, \,
\frac{\partial \psi}{\partial \nu_i} \approx \frac{g'_{\nu_i}}{\sqrt{w \overline{g^2}}} -
\frac{g\,\overline{g g'_{\nu_i}}}{\sqrt{w \overline{g^2}^3}},
\end{equation}
where the function $g$ and its derivatives $g'_x=\partial g/\partial x$ and
$g'_{\nu_i}=\partial g/\partial\nu_i$ outside the averaging are calculated for $x=2\pi
ft+\lambda$. This allows us to calculate the elements of the matrix $\mathbfss G$:
\begin{eqnarray}
\left\langle\left(\frac{\partial\psi}{\partial \lambda}\right)^2 \right\rangle &\approx& q \equiv
         \frac{\overline{{g'_x}^2}}{\overline{g^2}}, \nonumber\\
\left\langle\frac{\partial\psi}{\partial \lambda}
\frac{\partial\psi}{\partial \nu_i}\right\rangle &\approx& v_i \equiv
         \frac{\overline{g'_x g'_{\nu_i}}}{\overline{g^2}} -
         \frac{\overline{g g'_x}\,\overline{g g'_{\nu_i}}}{\overline{g^2}^2}, \nonumber\\
\left\langle \frac{\partial\psi}{\partial \nu_i}
\frac{\partial\psi}{\partial \nu_j}\right\rangle &\approx& V_{ij} \equiv
         \frac{\overline{g'_{\nu_i} g'_{\nu_j}}}{\overline{g^2}} -
         \frac{\overline{g g'_{\nu_i}}\,\overline{g g'_{\nu_j}}}{\overline{g^2}^2},
\label{vmv}
\end{eqnarray}
which do not depend on the frequency and phase. When calculating the elements of the
matrix $\mathbfss G$, we also deal with summations like, for instance,
\begin{equation}
\left\langle\left(\frac{\partial\psi}{\partial f}\right)^2\right\rangle \approx
         \frac{4\pi^2} {w \overline{g^2}} \left\langle t^2 {g'_x}^2(2\pi f t + \lambda) \right\rangle.
\label{Lsum}
\end{equation}
The $P$-periodic function ${g'_x}^2$ can be expanded in the Fourier series with the
constant term being equal to $\overline{{g'_x}^2}$. If $t_i$ span a large enough number of
the periods approximately uniformly, the summation in~(\ref{Lsum}) averages out all
Fourier harmonics, except for the constant term, which results in
\begin{equation}
\left\langle\left(\frac{\partial\psi}{\partial f}\right)^2\right\rangle \approx
         4\pi^2 \overline{t^2}\, q.
\end{equation}
Here, the line over $t^2$ denotes the weighted averaging of the squared timings:
$\overline{t^2} = \langle t^2 \rangle/\langle 1 \rangle$, which can be easily evaluated
directly (without approximating it by a continuous integral). Similar arguments lead to
\begin{equation}
\left\langle\frac{\partial\psi}{\partial f}
\frac{\partial\psi}{\partial \lambda}\right\rangle \approx
         2\pi \overline{t}\, q, \qquad
\left\langle\frac{\partial\psi}{\partial f}
\frac{\partial\psi}{\partial \nu_i}\right\rangle \approx
         2\pi \overline{t}\, v_i,
\end{equation}
where $\overline t$ is the weighted average of $t_i$. Thus, the full matrix $\mathbfss G$
can be written in the following block form:
\begin{equation}
\mathbfss G \approx \left(\begin{array}{ccc}
4\pi^2 \overline{t^2} q & 2\pi \overline{t} q & 2\pi \overline{t} \bmath v^{\rm T}\\
2\pi \overline{t} q & q & \bmath v^{\rm T} \\
2\pi \overline{t} \bmath v & \bmath v & \mathbfss V \\
\end{array}\right),
\label{matrixL}
\end{equation}
where the elements of the vector $\bmath v$ and those of the matrix $\mathbfss V$ are
defined in~(\ref{vmv}). In this approximation, the matrix $\mathbfss G$ only depends on
the parameters $\bnu$. Using simple elementary transformations of $\mathbfss G$ we can
finally obtain that
\begin{equation}
\det\mathbfss G \approx \pi q^2 T_{\rm eff}^2 \det\mathbfss R,
\label{detL}
\end{equation}
where $\mathbfss R = \mathbfss V - \bmath v \otimes \bmath v / q$ (that is, $R_{ij} =
V_{ij} - v_i v_j/q$) and $T_{\rm eff} = \sqrt{4\pi \left(\overline{t^2} -
\overline{t}^2\right)}$ is the effective length of the time series, as it was defined in
\citep{Baluev08a}. Integrating~(\ref{coefA}) over $\lambda$ and substituting~(\ref{detL}),
we can write down:
\begin{equation}
A \approx \frac{W}{\pi^{n/2-1}} \int\limits_\Upsilon q \sqrt{\det\mathbfss R}\, d\bnu,
\quad W = f_{\rm max} T_{\rm eff}.
\label{coefA-af}
\end{equation}
The factor $A$ can now be substituted in~(\ref{FAPhermite}) for the use in
(\ref{FAPbound}) and~(\ref{bound_fap}). In the degenerate case $n=2$, we have $\dim\bnu=0$
and put $\det\mathbfss R=1$ by definition.

The main advantage of this method of calculation of $A$ is that the result depends on a
particular time series only via the single quantity $T_{\rm eff}$. Given the model $g$ and
the domain $\Upsilon$, we can evaluate an approximation for $A$ once and then use it for
all time series, substituting the proper values of $f_{\rm max}$ and $T_{\rm eff}$. For
the LS periodogram, for instance, we have $A\approx W$, which is in full agreement with
\citep{Baluev08a}. The main disadvantage is that this approximation may have insufficient
accuracy in practice.

When deriving this approximation for $A$, we assumed the uniform phase coverage for all
frequencies $f$ in the scan range. When the original time series is not uniform, this
assumption may become invalid at some frequencies $f$, corresponding to periodic leakage
patterns of $t_i$ (and including also the zero frequency). However, these perturbations
can usually appear only inside very short frequency segments ($\Delta f \sim 1/T_{\rm
eff}$) around the leakage frequencies; for the most values of $f$ in the range $[0,f_{\rm
max}]$ the phase coverage is still uniform. But our formula for $A$ in~(\ref{coefA})
involves an integration over the wide frequency band $\Delta f \sim W/T_{\rm eff}$ with
$W\gg 1$, and hence the perturbations of its integrand have almost no effect on the
result, because they are limited by so short frequency segments. This means that the
accuracy of our approximation for $A$ does not significantly degrades even for non-uniform
time series. For sinusoidal signals this was already demonstrated in \citep{Baluev08a},
where we have shown that even ultimately strong aliasing has only a negligible effect on
the resulting $A$, when $W\gtrsim 10$.

In the case of non-sinusoidal signals, an important source of the errors of the
approximation~(\ref{coefA-af}) comes from another side. If the signal model contains some
quickly varying structures, e.g. narrow peaks, the observations may cover these structures
with insufficient sampling, so the resulting approximation for $A$ may appear poor even
when $t_i$ are perfectly uniform. This effect is important for the von Mises periodogram
below, for example.

\subsection{Evaluating $A$ directly}
\label{subsec_Adirect}
The direct evaluation of the factor $A$ by means of substitution of~(\ref{gradvar})
to~(\ref{coefA}) involves rather unpleasant manipulations with huge formulae, especially
when we work in general terms of Section~\ref{sec_nonlindef}. To simplify them, let us
write down the gradient of the model $h$ over the compound vector of all non-linear
parameters $\bomega=\{f,\lambda,\bnu\}$:
\begin{equation}
\bgamma = \frac{\partial h}{\partial \bomega} = \left\{2\pi t g'_x,\, g'_x,\, g'_\bnu \right\}.
\end{equation}
Then define
\begin{eqnarray}
\mathbfss Q_{\btheta_{\mathcal H},\btheta_{\mathcal H}} &=&
\langle \bvarphi_{\mathcal H} \otimes \bvarphi_{\mathcal H} \rangle, \nonumber\\
\mathbfss Q_{\btheta_{\mathcal H},K} &=& \langle \bvarphi_{\mathcal H}\, g \rangle, \nonumber\\
\mathbfss Q_{\btheta_{\mathcal H},\bomega} &=&
\langle \bvarphi_{\mathcal H} \otimes \bgamma \rangle, \nonumber\\
\mathbfss T &=& \langle \bgamma \otimes \bgamma \rangle -
                            \mathbfss Q_{\btheta_{\mathcal H},\bomega}^T
                            \mathbfss Q_{\btheta_{\mathcal H},\btheta_{\mathcal H}}^{-1}
                            \mathbfss Q_{\btheta_{\mathcal H},\bomega}, \nonumber\\
\bmath y &=& \langle g\, \bgamma \rangle -
                            \mathbfss Q_{\btheta_{\mathcal H},\bomega}^T
                            \mathbfss Q_{\btheta_{\mathcal H},\btheta_{\mathcal H}}^{-1}
                            \mathbfss Q_{\btheta_{\mathcal H},K}, \nonumber\\
D &=& \langle g^2 \rangle - \mathbfss Q_{\btheta_{\mathcal H},K}^T
                            \mathbfss Q_{\btheta_{\mathcal H},\btheta_{\mathcal H}}^{-1}
                            \mathbfss Q_{\btheta_{\mathcal H},K},
\end{eqnarray}
where $\bvarphi_{\mathcal H}$ is the functional base of the linear null model~(\ref{muH}).
Finally,
\begin{equation}
\mathbfss G = \frac{\mathbfss T}{D} - \frac{\bmath y \otimes \bmath y}{D^2}.
\label{Gdir}
\end{equation}

We may also offer another evaluation sequence. Let us construct the full Fisher
information matrix of all the parameters involved ($\btheta_{\mathcal H}$, $K$,
$\bomega$):
\begin{equation}
\mathbfss Q = \left\langle
\begin{array}{ccc}
\bvarphi_{\mathcal H} \otimes \bvarphi_{\mathcal H} & \bvarphi_{\mathcal H}\, g & \bvarphi_{\mathcal H} \otimes \bgamma \\
g\, \bvarphi_{\mathcal H}^{\rm T} & g^2 & g\, \bgamma^{\rm T} \\
\bgamma \otimes \bvarphi_{\mathcal H} & \bgamma\, g & \bgamma \otimes \bgamma
\end{array}
\right\rangle.
\label{fullFisher}
\end{equation}
Please notice that the triangle braces, standing for the weighted summation over the time
series, are still here. Now let us apply the Cholesky decomposition: $\mathbfss
Q=\mathbfss L \mathbfss L^T$ with $\mathbfss L$ being a lower-triangular matrix. Then
write down $\mathbfss L$ in the same block form as $\mathbfss Q$ in~(\ref{fullFisher}):
\begin{equation}
\mathbfss L = \left(
\begin{array}{ccc}
\mathbfss L_{\btheta_{\mathcal H},\btheta_{\mathcal H} } & 0 & 0 \\
\bmath l_{K,\btheta_{\mathcal H}}^{\rm T} & l_{K,K} & 0 \\
\mathbfss L_{\bomega,\btheta_{\mathcal H} } & \bmath l_{\bomega,K} & \mathbfss L_{\bomega,\bomega}
\end{array}
\right).
\end{equation}
Notice that $\bmath l_{K,\btheta_{\mathcal H}}$ and $\bmath l_{\bomega,K}$ are vectors,
and $l_{K,K}$ is a scalar. Obviously, $\mathbfss L_{\btheta_{\mathcal H},\btheta_{\mathcal
H} }$ is a lower-triangular matrix of the Cholesky decomposition for $\mathbfss
Q_{\btheta_{\mathcal H},\btheta_{\mathcal H} }$, and $\mathbfss L_{\bomega,\bomega}$ is
another lower-triangular matrix. From the matrix $\mathbfss L$ definition and
from~(\ref{Gdir}) we can easily derive two remarkable relations:
\begin{equation}
\mathbfss G = \frac{\mathbfss L_{\bomega,\bomega} \mathbfss L_{\bomega,\bomega}^{\rm T}}{l_{K,K}^2},
\qquad D=l_{K,K}^2.
\end{equation}
Moreover, it is clear that $\sqrt{\det\mathbfss G} = \det\mathbfss
L_{\bomega,\bomega}/l_{K,K}^d$. Therefore, to find the integrand in~(\ref{coefA}) we just
need to calculate $\det\mathbfss L_{\bomega,\bomega}$, which is simply equal to the
product of its diagonal elements, and then divide the result by $l_{K,K}^d$.

Therefore, the final procedure to evaluate $\sqrt{\det\mathbfss G}$ is as follows.
\begin{enumerate}
\item Evaluate $\mathbfss Q$ using its definition~(\ref{fullFisher}). Its size should be
$d_{\mathcal H}+d+1$. Note that it is important to preserve the ordering of the parameters
as $\btheta_{\mathcal H}, K, \bomega$, while the ordering inside $\btheta_{\mathcal H}$
and inside $\bomega$ is not important.
\item Perform the Cholesky decomposition of $\mathbfss Q$. It is very quick and numerically
stable procedure.
\item On the basis of only diagonal elements of the resulting Cholesky matrix $\mathbfss
L$, construct the combination $\left( \prod_{i=d_{\mathcal H}+2}^{d_{\mathcal H}+d+1}
l_{ii} \right) / \left( l_{d_{\mathcal H}+1,d_{\mathcal H}+1} \right)^d$. It is equal to
what we seek.
\end{enumerate}
The quantity $\sqrt{\det\mathbfss G}$ must be further numerically integrated over the
parameters $\bomega$, according to~(\ref{coefA}). Note that this includes the integration
over the frequency $f$ and over the phase $\lambda$ (which is now non-trivial).

Now, let us limit ourself to the most important practical case when the null model
involves only a free constant term: $d_{\mathcal H}=1$, $\varphi_{\mathcal H}\equiv 1$.
In this case the matrix $\mathbfss Q$ is considerably simplified:
\begin{equation}
\mathbfss Q = \left\langle
\begin{array}{ccc}
1 & g & \bgamma^{\rm T} \\
g & g^2 & g\, \bgamma^{\rm T} \\
\bgamma & g\, \bgamma & \bgamma \otimes \bgamma
\end{array}
\right\rangle.
\end{equation}
The round-off errors may destroy the positive-definiteness of $\mathbfss Q$, which is
critical for the Cholesky decomposition. To reduce this effect, the computational sequence
can be transformed to something similar to the formulae~(\ref{flm1}) and~(\ref{flm2}).
This will be some hybrid approach to evaluate $\sqrt{\det\mathbfss G}$ between the two
ones that we have already discussed. Namely, first we should center the functions
involved:
\begin{eqnarray}
g_c = \langle g \rangle / \langle 1 \rangle,
\quad \bgamma_c = \langle \bgamma \rangle / \langle 1 \rangle, \nonumber\\
\tilde g = g-g_c, \quad \tilde\bgamma = \bgamma-\bgamma_c.
\end{eqnarray}
After that, we need to evaluate
\begin{eqnarray}
D &=& \langle \tilde g^2 \rangle, \nonumber\\
\bmath y &=& \langle \tilde g\, \tilde\bgamma \rangle, \nonumber\\
\bbeta &=& \tilde\bgamma - \tilde g\, \bmath y/ D,
\end{eqnarray}
which imply that
\begin{equation}
\mathbfss G = \langle \bbeta \otimes \bbeta \rangle/D.
\end{equation}
We have no need to evaluate $\mathbfss G$ itself. Instead, we may perform the Cholesky
decomposition of $\langle\bbeta\otimes\bbeta\rangle$. Dividing the product of the diagonal
elements of the resulting Cholesky matrix by $D^{d/2}$, we obtain $\sqrt{\det\mathbfss
G}$.

The obious advantage of the direct method to evaluate the factor $A$ is that it allows for
high accuracy limited by only round-off and numerical integration errors, not relying on
any approximating assumptions. The disadvantage is that it is considerably more slow than
in Section~\ref{subsec_Aavg}, although in practice the relevant computation time should be
comparable to the time of a single evaluation of the periodogram itself.\footnote{We will
typically evaluate this periodogram on some multidimensional grid in the space of all the
parameters $\bomega$, not just on a frequency grid like in the LS case. In terms of the
computational demands, this procedure is roughly equivalent to the numerical integration
over the same space.} Therefore, this is still much faster than e.g. Monte Carlo
simulation, where this periodogram has to be re-evaluated thousands of times before we
reach a reliable $\FAP$ estimation.

\section{Unknown noise level}
\label{sec_likper}
In Section~\ref{sec_nonlindef}, we have assumed that the standard errors $\sigma_i$ of
observations are known a priori. In practice we often do not know them with enough
precision. A commonly used model is given by $\sigma_i^2 = \kappa/w_i$, where the
quantities $w_i$ determine the weighting pattern of the time series and the factor
$\kappa$ remains unconstrained a priori. Similar problem was considered in the paper
\citep{Baluev08b}. In this work, the model $\sigma_i^2 = \sigma_{{\rm meas},i}^2 +
\sigma_\star^2$ was considered with $\sigma_{{\rm meas},i}^2$ being the `internal'
measurements variances, known a priori, and parameter $\sigma_\star^2$ being the
unconstrained variance of the extra `jitter'. In these cases, we cannot calculate the
least-squares periodogram $z(f)$, since we cannot calculate the values of the $\chi^2$
functions themselves.

The general approach for solving such problems is based on the likelihood ratio test. The
logarithm of the likelihood function for our observations, which are contaminated by
random mutually independent Gaussian errors, may be written down (specifically for the
hypothesis $\mathcal H$ and $\mathcal K$) as
\begin{equation}
\ln \mathcal L_{\mathcal H,\mathcal K} = - \frac{1}{2} \sum_{i=1}^{N} \left[
\frac{(x_i-\mu_{\mathcal H,\mathcal K}(t_i))^2}{\sigma_i^2(\bmath p)} + \ln
\sigma_i^2(\bmath p) \right] + \const.
\end{equation}
Here the full variances $\sigma_i^2$ depend on the extra parameters $\bmath p$, which
should be estimated from the data together with the usual parameters $(\btheta_{\mathcal
H},\btheta,f)$ of the model curve. These estimations are obtained in result of maximizing
the corresponding likelihood function over all of the parameters to be estimated. After
that, we could construct the logarithm of the likelihood ratio statistic as the maximum
(over the frequency $f$) of the likelihood ratio periodogram
\begin{equation}
Z(f) = \max_{\bmath p,\btheta_{\mathcal K}} \ln\mathcal L_{\mathcal K}(\bmath
p,\btheta_{\mathcal K},f) - \max_{\bmath p,\btheta_{\mathcal H}} \ln\mathcal L_{\mathcal
H}(\bmath p,\btheta_{\mathcal H}).
\end{equation}
This function may give the basis for signal detection in the general framework, when
$\bmath p$ is not known a priori. However, for the aims of reduction of the statistical
bias in $\bmath p$, it is better to use the following modifications of the likelihood
functions and of the likelihood ratio periodogram:
\begin{eqnarray}
\ln \tilde{\mathcal L}_{\mathcal H,\mathcal K} = - \frac{1}{2} \sum_{i=1}^{N} \left[
\frac{(x_i-\mu_{\mathcal H,\mathcal K}(t_i))^2}{\gamma_{\mathcal H,\mathcal K}
\sigma_i^2(\bmath p)} + \ln \sigma_i^2(\bmath p) \right], \nonumber\\
\tilde Z(f) = \gamma_{\mathcal K} \left[ \max_{\bmath p, \btheta_{\mathcal K}}
\ln\tilde{\mathcal L}_{\mathcal K} - \max_{\bmath p, \btheta_{\mathcal H}}
\ln\tilde{\mathcal L}_{\mathcal H} \right] + \frac{N_{\mathcal K}}{2} \ln
\frac{N_{\mathcal H}}{N_{\mathcal K}},
\label{modlikrat}
\end{eqnarray}
where $N_{\mathcal H} = N - d_{\mathcal H}$, $N_{\mathcal K} = N - d_{\mathcal K}$, and
$\gamma_{\mathcal H,\mathcal K}=N_{\mathcal H,\mathcal K}/N$. This modification was
discussed in details in the paper \citep{Baluev08b}.

For the popular practical case $\sigma_i^2=\kappa/w_i$ with $w_i$ known a priori, the
periodogram $\tilde Z(f)$ represents a direct extension of the periodogram $z_3(f)$ from
\citep{Baluev08a}. It can be constructed now as the maximum (over $\btheta$) of the
non-linear function
\begin{eqnarray}
\zeta_3(\btheta,f) = - \frac{N_{\mathcal K}}{2} \ln\left[1 -
\frac{2\zeta(\btheta,f)}{\min_{\btheta_{\mathcal H}}\chi_{\mathcal H}^2(\btheta_{\mathcal
H})} \right].
\end{eqnarray}
Note that the ratio $\zeta/\chi_{\mathcal H}^2$ is independent of $\kappa$, so that
$\zeta_3$ can be calculated regardless the factor $\kappa$ is unknown.

To assess the statistical significance of the peaks on the likelihood ratio periodogram,
we need to know the distributions of the maxima of $\tilde Z(f)$, as previousl. Now we
cannot just apply the results from the previous sections directly. However, we may use the
asymptotic large-sample properties of the likelihood function.

To do this we have to assume that the new parameters $\bmath p$ do not introduce any extra
pecularities like e.g. extra non-identifiability under $\mathcal H$. This is normally
true. The calculations involving quadratic Taylor expansion of the likelihood function
near the point $K=0$, show that in the asymptotic $N\to\infty$ approximation for the
quantity $\eta$ looks exactly the same as its linear-case definition in~(\ref{eta})
and~(\ref{psi}). This means that the joint distribution of $\eta$ and of its gradient
remains asymptotically the same as in the genuine linear case. Concequently, all the
theory of Section~\ref{sec_FAP} remains valid for the likelihood-ratio periodograms as an
asymptotic approximation for $N\to\infty$. The same holds for the modified likelihood
functions and for the modified likelihood ratio periodogram~(\ref{modlikrat}), since this
modification does not introduce any change in the asymptotic behaviour.

For the periodogram $z_3(f)$ and linear models of the signal, the approximations to the
periodograms distributions for arbitrary $N$ are given in \citep{Baluev08a}. When
$N\to\infty$, these approximations indeed rapidly converge to those obtained for the
original least-squares periodogram. This provides an independent confirmation of the
arguments from the last paragraph. Therefore, for large datasets, we can apply the
analytic estimation of the $\FAP$ for the periodogram $z_3(f)$ and other non-linear
likelihood ratio periodograms in the same way as it was described in the previous sections
for least-squares periodograms with known noise uncertainties.

\section{Practical examples of non-linear periodograms}
\label{sec_examples}

\subsection{Detecting periodic signal with a fixed non-sinusoidal shape}
\label{sec_fixedshape}
Let us consider the case $d=2$ with $\btheta$ incorporating only the amplitude and phase
of the periodic signal to be detected. That is,
\begin{equation}
\mu = K g(2\pi f t + \lambda),
\end{equation}
where the $2\pi$-periodic function $g(x)$ is given a priori and is centred so that
$\overline{g}=0$. This function determines the shape of the putative periodic variation.

Notice that the term $M_{\rm boundary}$ in~(\ref{FAPbound}) may now only appear due to the
boundary points of the frequency segment. The phase $\lambda$ is a periodic parameter
defined over the self-closed circle which basically does not have a boundary. In other
words, all maxima of $\eta$ over $\lambda$ are local maxima where $\partial \eta/ \partial
\lambda =0$; no other maxima are possible in $\lambda$.

In this simple case we express $\FAP$ separately for the fixed-frequency and
unknown-frequency cases, using the approach of Section~\ref{subsec_Aavg}. In the fixed
frequency case, we find
\begin{equation}
\FAP_{\rm single} \lesssim M_{\rm single}(z) \approx \sqrt{q} e^{-z}.
\label{nonsinsingle}
\end{equation}

In the case of unknown frequency,
\begin{equation}
\FAP_{\rm max} \lesssim M_{\rm max}(z)+M_{\rm single}(z) \approx q W e^{-z} \sqrt{z} + \sqrt{q} e^{-z}.
\label{nonsinmax}
\end{equation}
The term $M_{\rm boundary}$ is equal to $M_{\rm single}$ here, since we have two end
points of the segment $[0,f_{\rm max}]$ and each should be counted as half.\footnote{This
is because there is a $50/50$ chance that such a boundary value is actually a boundary
minimim rather than a maximum, depending on the sign of the derivative in this end point.}
This term can be safely neglected in~(\ref{nonsinmax}) anyway.

Notice that as long as the approximation of the uniform phase coverage is valid, the
matrix $\mathbfss G$ appears here almost constant, so that we can use the simplified
formulae~(\ref{coefAconst}). This means that although in the right-hand side
of~(\ref{nonsinmax}) we omitted a term of the order of $e^{-z}/\sqrt z$, corresponding to
the term with $a_1$ of~(\ref{taunonlin}), the coefficient $a_1$ is itself negligibly
small. As we have discussed in \citep{Baluev08a} for the sinusoidal model, the
approximation of the uniform phase coverage works well even for time series with
ultimately strong spectral leakage. This is because the aliasing/leakage-induced errors
are concentrated only within a few narrow frequency segments, and they thus have only a
negligible effect on the quantities expressed by an integral over a large frequency band
(like $A$). However, for non-sinusoidal signals this approximation may appear poor due to
reasons unrelated to the spectral leakage effects; namely it may be poor when $g(x)$
demonstrates narrow peaks that our observations are unable to cover with enough dense
sampling.

In the general case $q\geq 1$. This inequality can be clearly derived by means of applying
the Parseval identity to the Fourier series for $g(x)$ and $g'(x)$. When $g$ is a harmonic
function, we deal with the LS periodogram. In this case, $\overline{g^2} =
\overline{{g'}^2} = 1/2$, and $q=1$. This allows us to entirely reproduce the exponential
single value distribution of the LS periodogram, $\FAP_{\rm single} = e^{-z}$, and the
Davies bound $\FAP_{\rm max} \lesssim W e^{-z} \sqrt{z} + e^{-z}$ from the paper
\citep{Baluev08a}. When $g(x)$ contains at least two Fourier terms we have $q>1$, so the
minimum $q=1$ is attained for the sinusoidal and only sinusoidal variation.

Just as a non-trivial example, let us consider the case when $g(x)$ has a sawtooth shape:
during the first half of its period it decreases linearly from $1$ to $-1$ and during the
second half it increases linearly from $-1$ to $1$. In this case, $\overline{{g'}^2} =
4/\pi^2$, $\overline{g^2} = 1/3$, and $q=12/\pi^2 \approx 1.216$.

\subsection{The von Mises periodogram}
\label{sec_vonmises}
Let us assume the following non-linear model for the periodic signal:
\begin{equation}
g(x,\nu) = \exp(\nu \cos x) - I_0(\nu), \qquad \nu>0.
\label{vonmises}
\end{equation}
In this definition $I_0$ stands for the modified Bessel function; we need it to satisfy
the condition $\overline g=0$. We can see that this function is $2\pi$-periodic in $x$. At
$\nu=0$ we have a formal singularity because $g(x,0)\equiv 0$ and~(\ref{psi}) becomes
degenerate. We can easily remove this degeneracy by making a replace
\begin{eqnarray}
\tilde g(x,\nu) = \frac{g(x,\nu)}{\nu} = \cos x+\frac{\nu}{2}\left(\cos^2 x - \frac{1}{2}\right) + \mathcal O(\nu^2), \nonumber\\
\tilde K = K\nu,
\label{smallnu}
\end{eqnarray}
so that for $\nu\to 0$ our model becomes equivalent to a simple sinusoid. For large $\nu$
the function~(\ref{vonmises}) represents a comb-like sequence of periodic narrow peaks,
each having the width $\sim 1/\sqrt\nu$. Note that a very similar function with a bit
different scaling, $\exp(\nu \cos x)/I_0(\nu)$, represents a probability density function
of the so-called von Mises distribution. This distribution is a periodic analog of the
Gaussian one, posessing a similar maximum-entropy property on a circle. For the sake of
convinience we will also call~(\ref{vonmises}) as the von Mises function, since these
small differences in the centering and normalization are not very important for us.

\begin{figure}
\includegraphics[width=84mm]{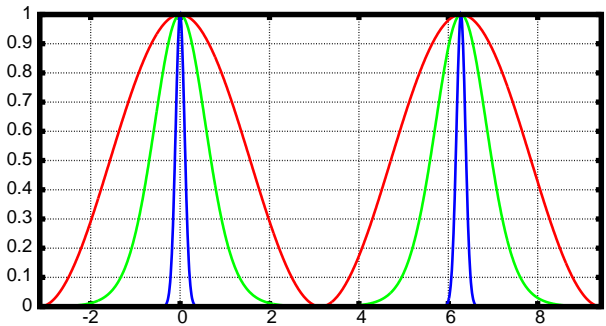}
\caption{The graphs of the von Mises function $g(x,\nu)$ for $\nu=0$ (the sinusoid),
$\nu=3$, and $\nu=100$ (the most peaky case). All plots are prescaled to always cover the
range $[0,1]$ in the abscissa.}
\label{fig_vonmises}
\end{figure}

In Fig.~\ref{fig_vonmises}, we plot the von Mises function for several values of the
localization parameter $\nu$. Looking at these plots, we might notice that such shape may
provide a satisfactory generic approximation to many physical variabilities that emerge in
the astronomical practice. In particular, it may appear good for the lightcurves of many
variable stars and planetary transits (just turn these graphs upside-down to optain
something similar to a transit). Since such model is functionally very simple (and thus
quickly calculatable and easy in various analytic manipulations), it looks rather tempting
to construct a periodogram that could utilise it as a model of the probe periodic signal.

Assume that we scan this periodogram in a rectangle $\nu_{\rm min} \leq \nu \leq \nu_{\rm
max}$ and $0 \leq f \leq f_{\rm max}$, and disallowing the signal amplitude $K$ to become
negative. Then we should first evaluate the function $\tau$ using~(\ref{taunonlin}) for
the interior of this rectangle (implying $n=3$). It will be approximately proportional to
$W$. Also, we should evaluate the function $M_{\rm boundary}$, which contains the term
responsible for the four sides and four vertices of the mentioned rectangle. Among these,
only the terms due to the sides $\nu=\nu_{\min,\max}$ are important. This is becuase these
are the only boundary terms proportional to $W$. The boundary terms due to the two other
sides and due to the vertices do not contain this multiplier, because the frequency $f$ is
held fixed there. Since in practice $W$ is large or very large, we only need to take into
account the boundary edges running along the frequency axis.

The final approximation to the false alarm probability can be represented in the form
\begin{eqnarray}
\FAP(z) \lesssim M(z) = W e^{-z} \left[ (X(\nu_{\rm max})-X(\nu_{\rm min})) z + \phantom{\frac{\sqrt z}{2}} \right. \nonumber\\
                            \left. + (Y(\nu_{\rm min})+Y(\nu_{\rm max})) \frac{\sqrt z}{2} + \mathcal O(z^0) \right],
\label{FAPvm}
\end{eqnarray}
where
\begin{eqnarray}
X(\nu) = \frac{1}{2 W\pi^2} \int\limits_0^{f_{\rm max}} df \int\limits_0^\nu d\nu
\int\limits_0^{2\pi} \sqrt{\det\mathbfss G_{f\lambda\nu}(f,\lambda,\nu)}\, d\lambda, \nonumber\\
Y(\nu) = \frac{1}{2 W\pi^{3/2}} \int\limits_0^{f_{\rm max}} df
\int\limits_0^{2\pi} \sqrt{\det\mathbfss G_{f\lambda}(f,\lambda,\nu)}\, d\lambda.
\label{coeffsXY}
\end{eqnarray}

In~(\ref{FAPvm}), the terms involving the factor $X$ correspond to the local maxima of
$\eta$ in the interior of the rectangle. The difference $X(\nu_{\rm max})-X(\nu_{\rm
min})$ is because the factor $A$ now contains an integral from $\nu_{\rm min}$ to
$\nu_{\rm max}$, while the function $X(\nu)$ is defined as an integral from $0$ to $\nu$.
The terms with $Y$ are for maxima on the two boundary lines $\nu=\nu_{\rm min}$ and
$\nu=\nu_{\rm max}$. The sum $Y(\nu_{\rm min})+Y(\nu_{\rm max})$ is because we need to sum
the maxima at the both borders, and the extra multiplier of $1/2$ is because we should
filter out half of these boundary maxima, due to the derivative $\partial\eta/\partial\nu$
having at these maxima an inappropriate sign with probability $1/2$. The quantities $W
X(\nu)$ and $W Y(\nu)$ represent, in fact, the factor $A$ for the rectangle $[0, \nu]
\times [0,f_{\rm max}]$ and for the boundary segments $\{\nu=\nu_{\rm min},\nu_{\rm max}\}
\times [0,f_{\rm max}]$.

In~(\ref{coeffsXY}), the $3\times 3$ matrix $\mathbfss G_{f\lambda\nu}$ corresponds to the
full gradient of $\eta$ over all three non-linear parameters $f,\lambda,\nu$; the matrix
$\mathbfss G_{f\lambda}$ is a $2\times 2$ submatrix of $\mathbfss G_{f\lambda\nu}$
involving only the elements corresponding to only the parameters $f$ and $\lambda$. Both
these matrices depend on all three parameters. Notice that the terms in~(\ref{FAPvm})
containing $Y$ basically correspond to the signal having a fixed non-sinusoidal shape
($\nu$ is fixed), and thus they can be also treated using the formalism of
Sect.~\ref{sec_fixedshape}.

In the particular case $\nu_{\rm min}=0$ (with the sign of $K$ still fixed) we should take
into account the obvious equality $X(0)=0$:
\begin{eqnarray}
\FAP(z) \lesssim M(z) = W e^{-z} \left[ z X(\nu_{\rm max}) + \phantom{\frac{\sqrt z}{2}} \right .\nonumber\\
              \left. + (Y(\nu_{\rm max})+Y(0)) \frac{\sqrt z}{2} + \mathcal O(z^0) \right].
\label{FAPvm_zero}
\end{eqnarray}
Notice that in practice $Y(0)\approx 1$ with good precision (see below).

When $K$ is allowed to be positive as well as negative we should double the right-hand
side of the expression in~(\ref{FAPvm}). This is because we should now honour the local
minima of the random field $\eta$ too, as well as its local maxima. We have in this case:
\begin{eqnarray}
\FAP(z) \lesssim M(z) = W e^{-z} \left[ 2 (X(\nu_{\rm max})-X(\nu_{\rm min})) z + \phantom{\sqrt z} \right. \nonumber\\
                            \left. + (Y(\nu_{\rm min})+Y(\nu_{\rm max})) \sqrt z + \mathcal O(z^0) \right],
\end{eqnarray}

A special case occures when $\nu_{\rm min}=0$ and the sign of $K$ is arbitrary. Then we
have, basically, some degeneracy of the free variables at $\nu=0$, making the cases $K<0$
and $K>0$ equivalent to each other (due to the symmetry of the sinusoid). This property
can be used to refine the Rice bound a bit. Assume that we have some point at the boundary
$\nu=0$, such that $\eta>0$ (implying $K>0$) and $\partial\eta/\partial\nu<0$. It is easy
to show that at a dual point with $\lambda \mapsto \lambda+\pi$ the value of $\eta$
changes the sign (hence $K<0$), but the value of $\partial\eta/\partial\nu$ remains
exactly the same. This is because the derivative of $\tilde g$ in~(\ref{smallnu}) over
$\nu$ is a $\pi$-periodic function of $x$ (at $\nu=0$), while the model $\tilde g$ itself
is $2\pi$-periodic. Therefore, the derivative $\partial|\eta|/\partial\nu$ has different
sign in these dual points, while the value of $|\eta|$ is identical.

Therefore, although there are many ``good'' boundary maxima at $\nu=0$, which satisfy the
condition $\partial|\eta|/\partial\nu<0$ (meaning that $|\eta|$ necessarily decreases when
we step from $\nu=0$ inwards), each such a maximum has a dual ``bad'' maximum at
$\lambda+\pi$, where $\partial|\eta|/\partial\nu>0$. Since there are no constraints on
$\lambda$, this means that for \emph{any} boundary maximum at $\nu=0$, either ``good'' or
``bad'', we can find larger values of $\eta$ in the interior $\nu>0$. This implies that
the \emph{global} maximum of $|\eta|$ cannot be attained at the line $\nu=0$. Following
the terminology by \citet{AzaisDelmas02} (see their Theorem~2), we basically proved that
the field $\zeta$ (or $|\eta|$) is a ``field without boundary'', concerning the boundary
line $\nu=0$. This allows us to just drop the relevant boundary term with $Y(\nu_{\rm
min}=0)$:
\begin{equation}
\FAP(z) \lesssim \tilde M(z) = W e^{-z} \left[ 2z X(\nu_{\rm max}) + Y(\nu_{\rm max}) \sqrt z + \mathcal O(z^0) \right],
\end{equation}
Notice that it is essential here that $K$ has arbitrary sign, because otherwise we could
not freely swap the values $\eta>0$ with $\eta<0$.

Let us first apply the approach of the Section~\ref{subsec_Aavg} to evaluate $X$ and $Y$.
The derivatives of the function~(\ref{vonmises}) look like
\begin{eqnarray}
g'_x   &=& -\nu \exp(\nu \cos x) \sin x, \nonumber\\
g'_\nu &=& \exp(\nu \cos x) \cos x - I_1(\nu).
\label{vonmises_der}
\end{eqnarray}
Substituting them to~(\ref{vmv}), and using the well-known integral representations for
the modified Bessel functions $I_k$, we obtain
\begin{eqnarray}
\overline{g^2} = I_0(2\nu) - I_0^2(\nu), \quad \overline{{g'_x}^2} = \frac{\nu^2}{2} \left[I_0(2\nu)-I_2(2\nu)\right], \nonumber\\
\overline{{g'_\nu}^2} = \frac{1}{2} \left[ I_0(2\nu) + I_2(2\nu) \right] - I_1^2(\nu), \qquad \overline{g'_x g'_\nu} = 0, \nonumber\\
\overline{g g'_\nu} = I_1(2\nu) - I_0(\nu) I_1(\nu),
\label{vmv_vonmises}
\end{eqnarray}
and then
\begin{eqnarray}
q = \frac{\nu^2}{2} \frac{I_0(2\nu)-I_2(2\nu)}{I_0(2\nu) - I_0^2(\nu)}, \qquad v=0, \nonumber\\
\det \mathbfss R = V = \frac{I_0(2\nu) + I_2(2\nu) - 2 I_1^2(\nu)}{2 [I_0(2\nu) - I_0^2(\nu)]} - \nonumber\\
    - \left[\frac{I_1(2\nu) - I_0(\nu) I_1(\nu)}{I_0(2\nu) - I_0^2(\nu)}\right]^2.
\end{eqnarray}
The behaviour of these quantities is not obvious from these formulae, so we need to
understand their asymptotic behaviour. For small $\nu$ we can use the Tailor expansion of
$I_k(z)$ to find that $q(0)=1$ and $V(0)=1/16$. This implies that for the factor $X$ the
integrand $q\sqrt{\det\mathbfss R}$ is equal to $1/4$ at $\nu=0$; for the factor $Y$ we
have $\det \mathbfss R=1$ by definition and $q\sqrt{\det\mathbfss R}$ at $\nu=0$ is unit.
For large $\nu$ we may use the following asymptotically converging expansion:
\begin{equation}
I_k(z) \simeq \frac{e^z}{\sqrt{2\pi z}} \left(
 1 - \frac{4k^2-1}{8z} + \frac{(4k^2-1)(4k^2-9)}{2! (8z)^2} + \ldots \right),
\end{equation}
which can be found e.g. in \citep[\S 21.8]{Korn}. The calculations lead us to
\begin{eqnarray}
q \simeq \frac{\nu}{2}, \qquad V \simeq \frac{1}{8\nu^2}, \nonumber\\
q \sqrt{\det\mathbfss R} \simeq \frac{1}{4 \sqrt 2} \quad ({\rm for\ }X), \qquad
q \sqrt{\det\mathbfss R} \simeq \frac{\nu}{2} \quad ({\rm for\ }Y).
\end{eqnarray}
When calculating $X$, we should further integrate the relevant quantity over $\nu$, so we
obtain $X(\nu)\sim \nu$ for large $\nu$. The factor $Y$ does not request this integration,
but its asymptotics appears eventually the same, $Y(\nu)\sim \nu$ for large $\nu$. Also,
for large $\nu$ we have $Y(\nu)/X(\nu)\simeq 2\sqrt{2\pi}$, hence the ``primary'' $X$-term
in~(\ref{FAPvm}) really exceeds the $Y$-term only for $z>2\pi$, and even beyond this level
they remain mutually comparable up to rather large $z$. This means that both these terms
should be taken into accound in practice; none can be neglected.

\begin{figure}
\includegraphics[width=84mm]{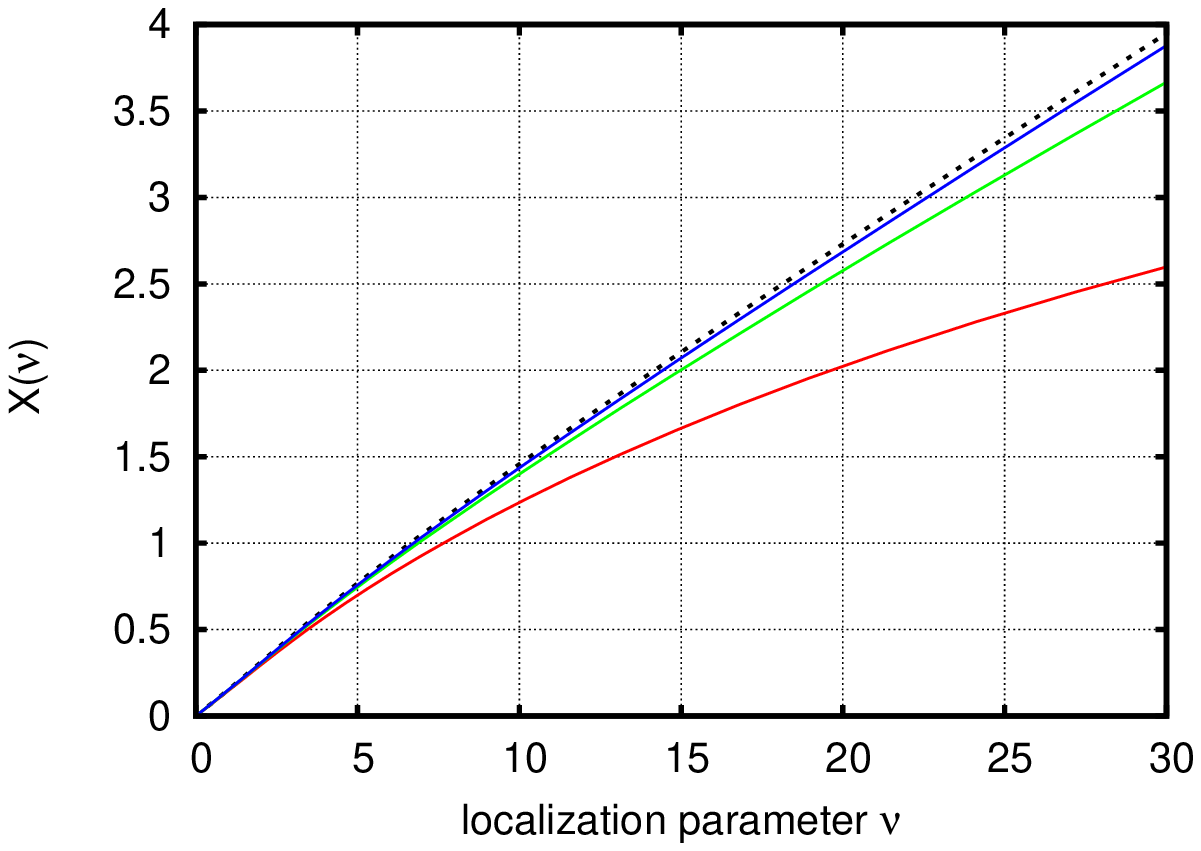}\\
\includegraphics[width=84mm]{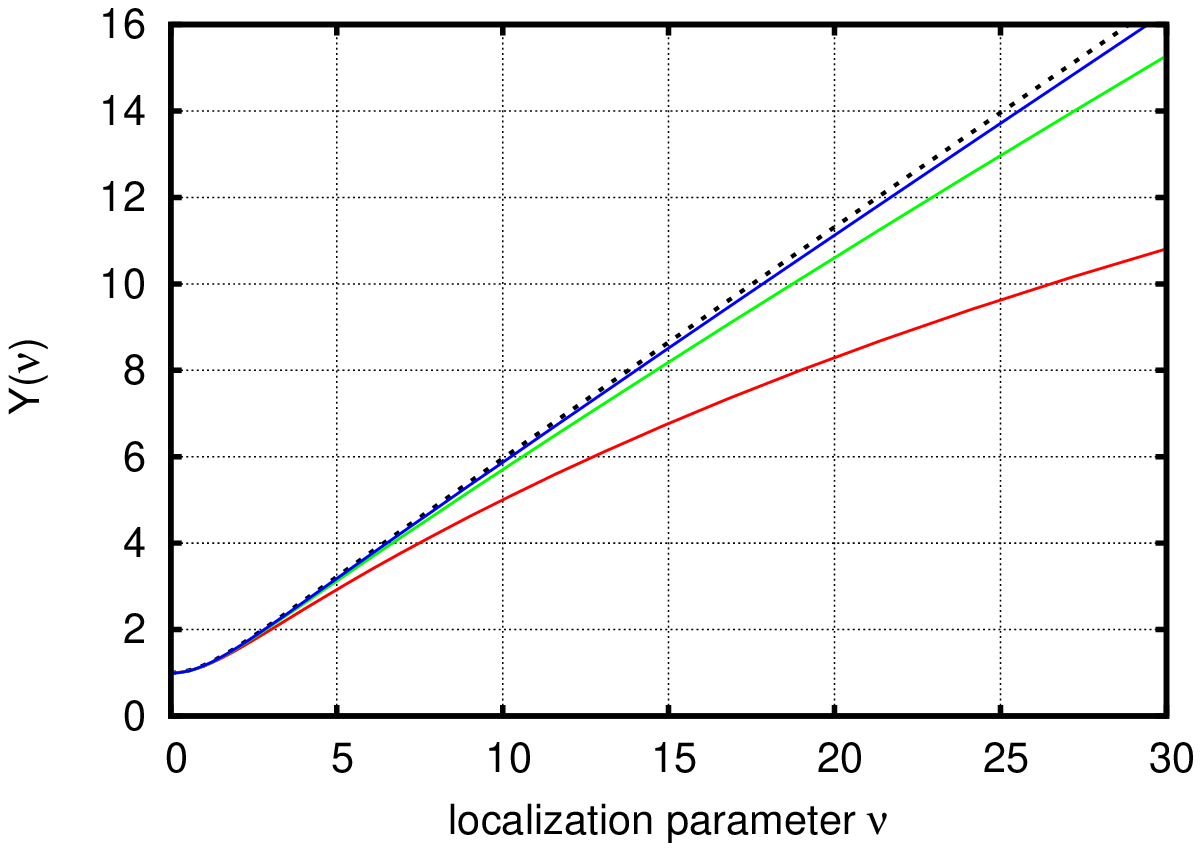}
\caption{The coefficients $X$ and $Y$ for the von Mises periodogram. Dashed curve is for
the approximation of Section~\ref{subsec_Aavg}, while the three solid curves correspond to
the direct precise method of Section~\ref{subsec_Adirect}. These three curves (from down
to up) correspond to three simulated time series, containing $N=30$, $100$, and $1000$
randomly distributed simulated observations. For all cases we have $W\approx 100$.}
\label{fig_XY}
\end{figure}

These also results infer that the integral in~(\ref{coefA-af}) is infinite if we do not
limit $\nu$ from the upper side. In this regard the parameter $\nu$ is similar to the
frequency $f$. When we scan the usual periodogram in a wider frequency range, the
probability to find a high noisy peak in this range inevitably increases. Similarly, an
attempt to detect more quickly-varying signals with larger $\nu$ will inevitably increase
our chances to catch a noisy fluctuation instead of a true signal.

We find that the approximate expressions for the factors $X$ and $Y$ obtained using the
formalism of Section~\ref{subsec_Aavg} a very accurate for small $\nu$, but this accuracy
decreases when $\nu$ grows, and increases when $N$ grows (Fig.~\ref{fig_XY}). We assume
that the error of this approximation emerges because our $N$ observations cannot sample
well the narrow peaks of the signal having the width $\sim 1/\sqrt \nu$.
\begin{figure*}
\begin{tabular}{@{}c@{}c@{}}
\includegraphics[width=0.49\textwidth]{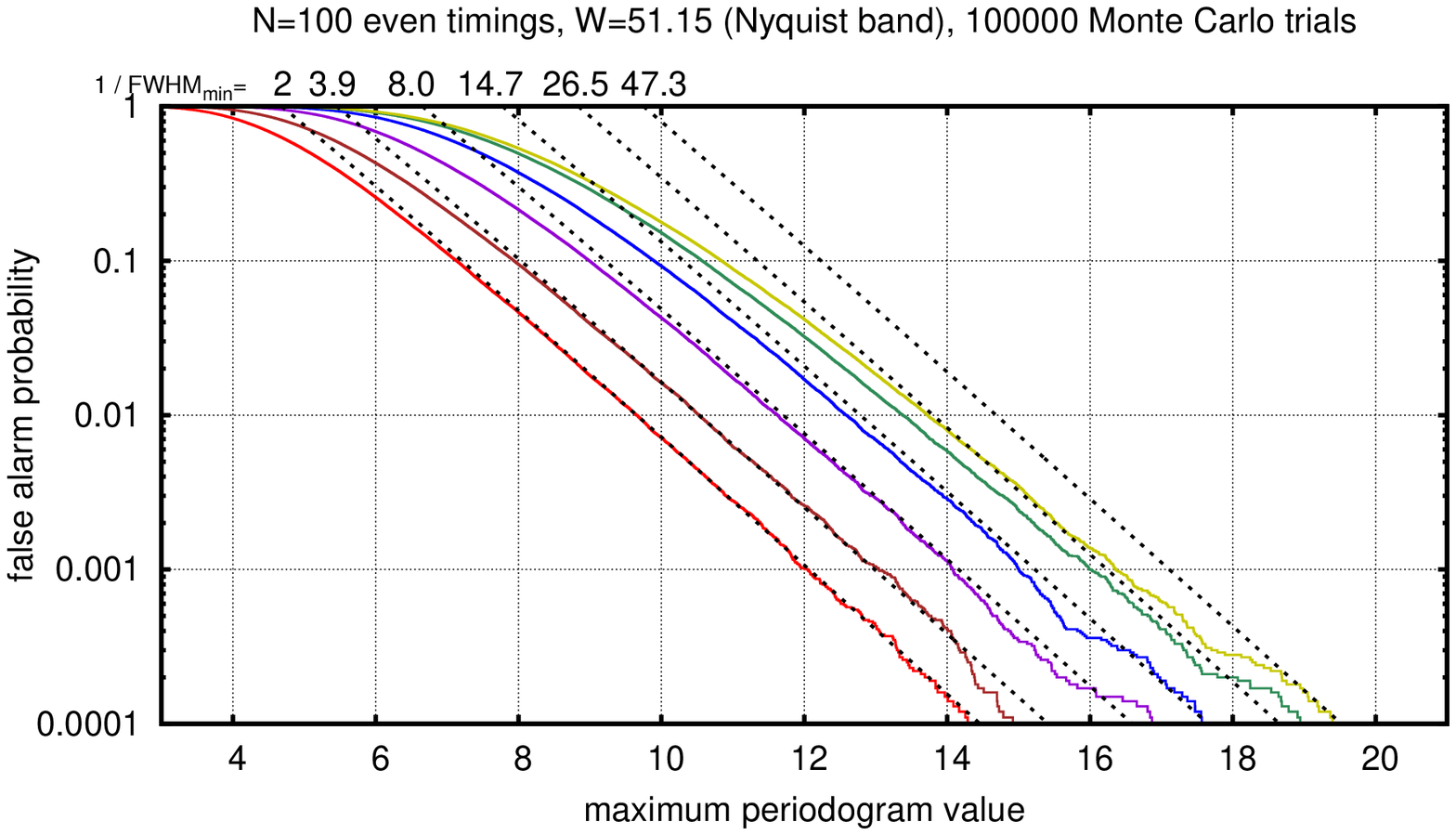} &
\includegraphics[width=0.49\textwidth]{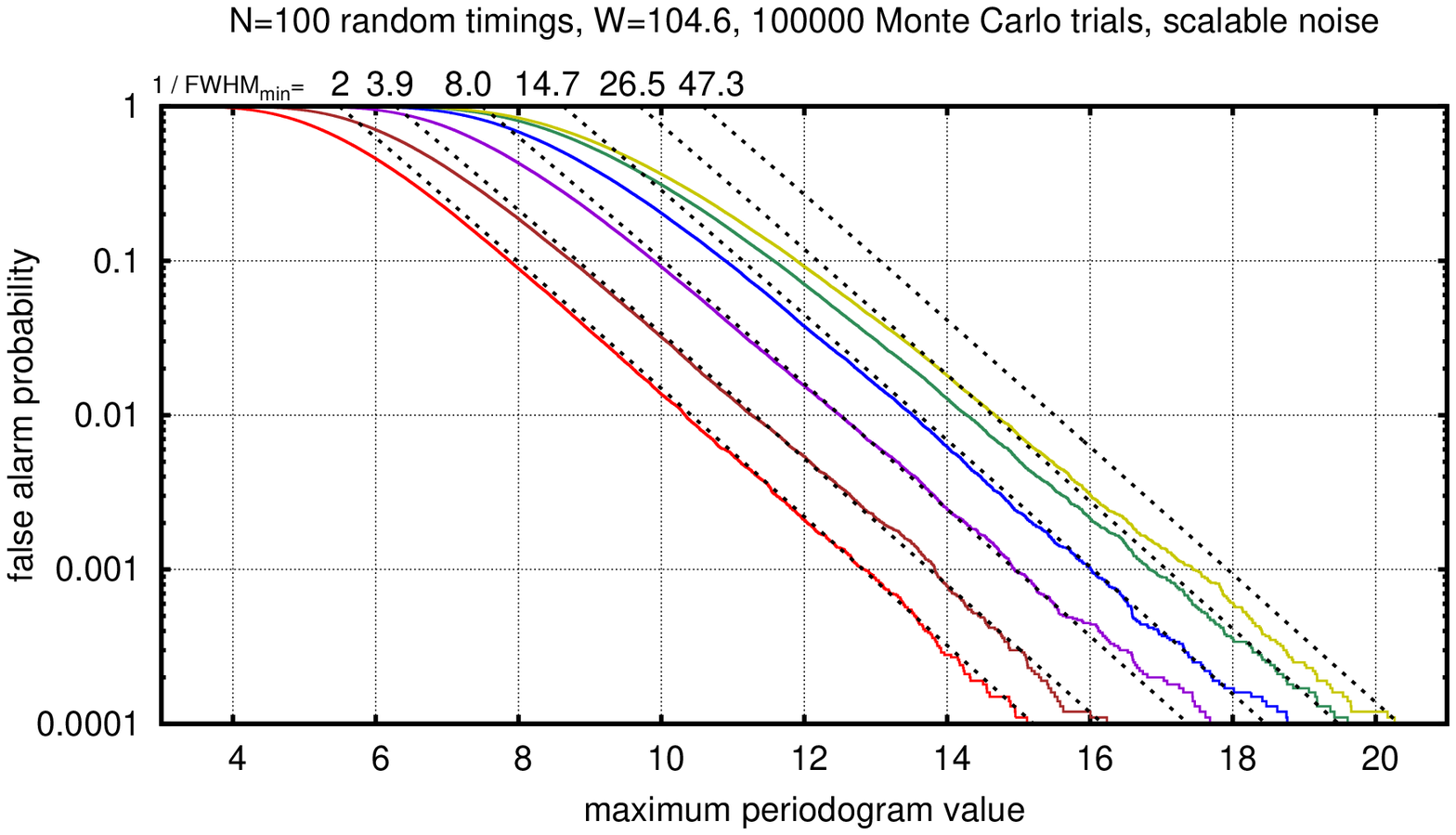} \\
\includegraphics[width=0.49\textwidth]{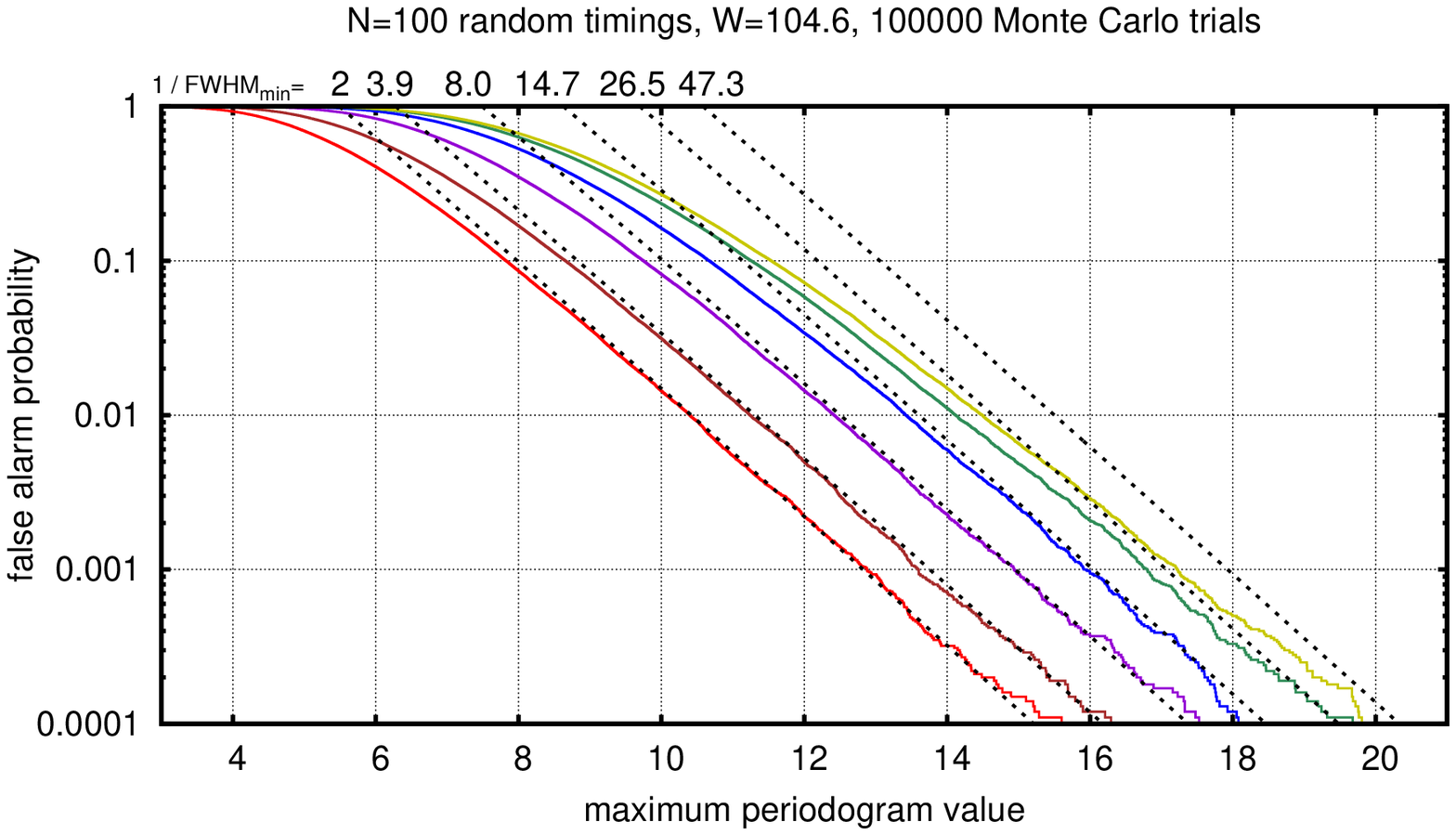} &
\includegraphics[width=0.49\textwidth]{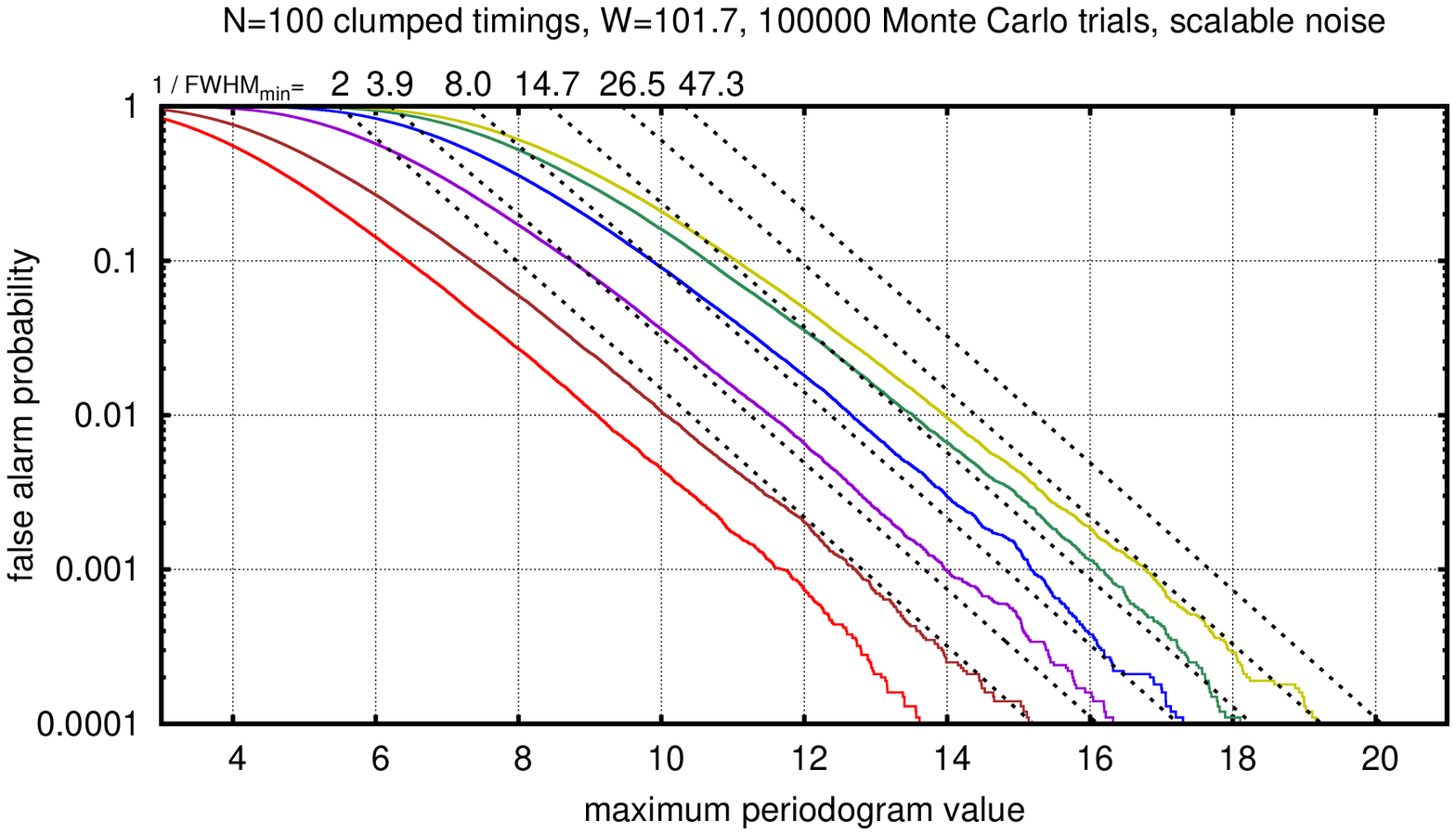}
\end{tabular}
\caption{The $\FAP$ approximation of the von Mises periodogram in comparison with Monte
Carlo simulations. In each panel we show the simulated and analytic $\FAP$ curves for
$\nu_{\rm min}=0$ and six different values of $\nu_{\rm max}$ from $0$ to $\approx 315$.
In the graphs we mark, instead of the upper limit for $\nu$, a lower limit for a more
intuitive FWHM (Full Width at Half Maximum) characteristic of the signal peaks. It can be
easily mapped one-to-one with $\nu_{\rm max}$ and it varies here from $1/2$ (corresponding
to the sinusoidal variation) to approximately $1/50$ (typical for e.g. a planetary
transit). The panels to the left correspond to the cases with the noise uncertainties
$\sigma_i$ known a priori; the ones to the right are for the multiplicative noise model of
Sect.~\ref{sec_likper}, $\sigma_i^2 \propto 1/w_i$. In the case of ``clumped random
timings'' (right-bottom panel) the $N=100$ points of the time series were equally split in
$10$ equidistant groups with $90\%$ gaps between them. This implies a very strong
aliasing, which makes our analytic approximation for $\FAP$ relatively inaccurate, though
they still work as an upper limit.}
\label{fig_distr_vM}
\end{figure*}

We have done some Monte Carlo simulations to test the quality of the
approximation~(\ref{FAPvm_zero}). The results are shown in Fig.~\ref{fig_distr_vM}. We
find that our analytic approximations behave exactly as we might expect. They have good
accuracy for practically important small levels of the $\FAP$ and for not too large
$\nu_{\rm max}$ and time-series leakage. For larger $\nu_{\rm max}$ or for strong leakage
the accuracy somewhat degrades, but the formula~(\ref{FAPvm_zero}) still works as an upper
bound on $\FAP$. Our conclusion is that this formula would be certainly useful in
practical applications.

\begin{figure}
\includegraphics[width=84mm]{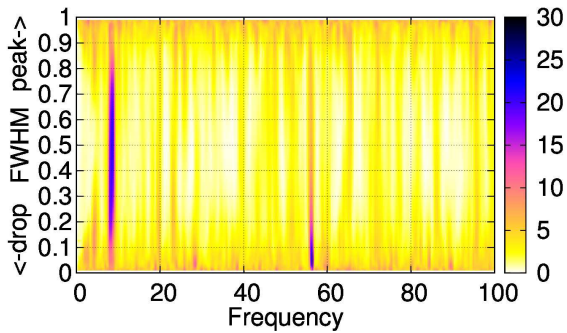}
\caption{The von Mises periodogram plotted for a test time series containing two periodic
signals contaminated by noise (see text). The lower half (${\rm FWHM}<0.5$) of the graph
correspond to a drop-type periodicity, while the upper part (for which we made a symbolic
replacement ${\rm FWHM}\mapsto 1-{\rm FWHM}$ for convenience, so the labelled values ${\rm
FWHM}$ exceed $0.5$) is for a peak-type variation. The slice at ${\rm FWHM}=0.5$
represents the LS periodogram assuming the sinusoidal model of the signal.}
\label{fig_test}
\end{figure}

At last we would like to demonstrate the power of the von Mises periodogram itself. We
generated a simulated time series with $N=1000$ randomly spaced observations. The values
of the simulated measurements contained two periodic signals with comparable amplitudes: a
sinusoidal variation and periodic flat drops (simulating planetary transits). Both signals
were below the noise level, so the noise should provide significant contamination. The von
Mises periodogram of these data is plotted in Fig.~\ref{fig_test}. We can see that it
allows an easy detection of the both signals at once ($f\approx 8$ and $f\approx 56$),
while the LS periodogram (wich represents, basically the middle horizontal slice of the
plot) would robustly reveal only the sinusoidal periodicity, allowing the planetary
transit to slip away until the next step of the analysis.

\section{Conclusions}
In this paper, we extended our previous results \citep{Baluev08a,Baluev09a} to the case
when the model of the signal to be detected in the noisy data depends on unknown
parameters in a non-linear manner. The definition of the periodogram was extended to this
non-linear (and non-sinusoidal) case in terms of the $\chi^2$ and likelihood-ratio tests.
We described a generic method of constructing an asymptotic approximation to the false
alarm probability. Based on these general results, we considered two specialized
non-linear periodograms. The first one involves a fixed-shape periodic non-sinusoidal
model of the signal. The second one models the signal with the so-called von Mises
function $\exp(\nu\cos x)$. This function is very remarkable, because it allows fairly
good approximation of very different periodic variations, from the plain sinusoid to a
model with periodic narrow peaks or drops (typical for e.g. the exoplanetary transit
lightcurve). For both these periodograms we provide a complete theoretical solution of the
false alarm probability problem.

Moreover, for the von Mises periodogram we offer a supporting package of C++ programs,
that may dramatically faciliate the use of the relevant theory in practice. This package
is attached to the present paper as the online-only supporting material (as a compressed
archive).

We expect that the results of this work can be used in a wide variety of astronomical
applications that deal with non-sinusoidal periodicities in observational data. These
research fields are ranged from the studies of variable stars to the studies of extrasolar
planetary systems.

In the forthcoming and future work, we plan to apply our approach to the so-called
double-frequency periodogram, where the signal is modelled by a sum of two independent
sinusoidal terms (to appear in Astron. Lett., under review), to the Schuster periodogram
and to the so-called Keplerian periodogram introduced by \citet{Cumming04}, also known as
``2DKLS periodogram'' \citep{OToole09b}.

\section*{Acknowledgments}
This work was supported by the Russian Foundation for Basic Research (project 12-02-31119
mol\_a) and by the Programme of the Presidium of Russian Academy of Sciences
``Non-stationary phenomena in the objects of the Universe''. I would like to express my
gratitude to the anonymous reviewer, who provided suggestions of a great value.

\bibliographystyle{mn2e}
\bibliography{nonlin}

\bsp

\label{lastpage}

\end{document}